\documentclass[aps,pra,reprint,showpacs,showkeys,floatfix]{revtex4-1}

\usepackage{graphicx}
\usepackage{lipsum}
\usepackage{color}
\usepackage{xcolor}
\usepackage{amssymb}
\usepackage{amsmath}
\usepackage{epsfig}
\usepackage{bm}
\usepackage[american]{babel}
\usepackage{hyperref}
\usepackage{braket}
\usepackage{tabularx}
\usepackage{wasysym}
\usepackage[T1]{fontenc} 
\usepackage{txfonts} 
\usepackage{comment}
 
\DeclareGraphicsExtensions{.pdf,.png,.jpg}
\hypersetup{colorlinks=true,urlcolor=blue,breaklinks=true,citecolor=blue,linkcolor=blue,pdfstartview=FitH,pdfpagemode=UseNone}

\begin{document}

\title{Fidelity measurement of a multiqubit cluster state with minimal effort}

\author{Konstantin Tiurev}
\author{Anders S. S{\o}rensen}
\affiliation{Center for Hybrid Quantum Networks (Hy-Q), The Niels Bohr Institute, University of Copenhagen, DK-2100 Copenhagen {\O}, Denmark}

\begin{abstract}
The size of the Hilbert space for a multiqubit state scales exponentially with the number of constituent qubits. Often this  leads to a similar exponential  scaling of the experimental resources  required to  characterize the state. 
Contrary to this, we propose a physically-motivated method for experimentally assessing the fidelity of an important class of entangled states known as  cluster states. The proposed method always yields a lower bound of the fidelity with  a number of measurement settings scaling only linearly with the system size, and is tailored to correctly account for the errors most likely to occur in experiments. For one-dimensional cluster states, the constructed fidelity measure is tight to lowest order in the error probability for experimentally realistic noise sources and thus closely matches the true fidelity. Furthermore, it is tight for the majority of higher-order errors, except for a small subset of certain non-local multiqubit errors irrelevant in typical experimental situations. The scheme also performs very well for higher-dimensional cluster states, assessing  correctly the majority of experimentally relevant errors. 
\end{abstract}

\maketitle

\section{Introduction}
With the tremendous progress in quantum technologies, the number of entangled qubits in quantum devices is rapidly increasing~\cite{PhysRevLett.106.130506,PhysRevLett.121.250505,PhysRevLett.122.110501}. 
In such devices, multiqubit entangled states can be utilized as essential resources for quantum information processing tasks, including quantum computation~\cite{RevModPhys.79.135,PhysRevLett.93.040503,PhysRevLett.95.010501,Knill:2001aa,doi:10.1063/1.5115814,doi:10.1063/1.4976737} and error correction~\cite{RevModPhys.87.307}, quantum communication~\cite{Azuma:2015aa,Li:2019aa,Buterakos2017,hilaire2020resource,PhysRevX.10.021071,PhysRevA.95.012304}, quantum simulations~\cite{Lanyon:2010aa,Ma:2011aa}, and cryptographic protocols~\cite{PhysRevLett.98.020503,PhysRevLett.95.200502,PhysRevA.69.052307,PhysRevA.72.044301}. These applications require reliable and robust preparation of the desired state. Accurate characterization of the multiqubit states produced in experiments is therefore essential. 

The standard workhorse for state verification is quantum state tomography~(QST)~\cite{PhysRevA.40.2847,PhysRevA.64.052312,PhysRevA.66.012303,RevModPhys.81.299,Banaszek_2013,ALTEPETER2005105}, enabling  full reconstruction of the density matrix. This procedure has been successfully utilized for few-qubits states~\cite{10.1038/nature04279,10.1038/nphys507,PhysRevLett.74.884,PhysRevLett.95.210502,10.1038/nphys507,PhysRevLett.117.210502,PhysRevLett.121.250505}, but becomes practically impossible for larger systems because the number of measurements settings, amount of measurement time, and computational resources required to complete the tomography grow exponentially with the number of particles. These protocols become even more impractical in situations where switching between measurement settings is costly or time consuming. To this end, considerable effort has been dedicated to devising experimentally efficient and practical tools for quantum state verification~\cite{PhysRevLett.105.150401,PhysRevLett.105.250403,Cramer2010,Lanyon2017,PhysRevLett.106.230501,PhysRevLett.107.210404,PhysRevLett.111.020401,PhysRevLett.105.250403}. Various resource-efficient tomography schemes have been proposed in recent years~\cite{PhysRevLett.105.250403,Lanyon2017,Cramer2010}. While these schemes scale more favourably than the full QST, state verification is still a daunting task in terms of measurement settings and computational postprocessing of the collected data~\cite{PhysRevLett.105.250403,Lanyon2017,Cramer2010}, or requires additional demanding control capabilities such as unitaries on the subsystems~\cite{Cramer2010}. Other elaborate approaches to state characterization utilize adaptive~\cite{PhysRevLett.113.190404,PhysRevLett.117.040402,Qi2017} or machine learning-assisted~\cite{Lohani_2020,Torlai2018b} quantum state tomography. Although reducing  the amount of hardware required for state characterization, it still presents an intractable task as the system size grows. 

For certain states, state verification can be optimized based on symmetries of the states. So called  Greenberger--Horne--Zeilinger and W states allow for efficient fidelity evaluation with resources increasing linearly in the system size~\cite{PhysRevA.76.030305}. Among other classes of entangled states, the so-called cluster states are of particular interest since they provide a means for universal quantum computation solely by measurements~\cite{RevModPhys.79.135,PhysRevLett.93.040503,PhysRevLett.95.010501,PhysRevLett.95.010501,Knill:2001aa,Azuma:2015aa,Li:2019aa,PhysRevX.10.021071}. A simple lower bound on the fidelity of  cluster states can be obtained using  only two measurement settings via direct measurement~\cite{PhysRevLett.117.210504,PhysRevLett.94.060501,GUHNE20091,PhysRevA.76.030305,PhysRevA.72.022340,PhysRevLett.103.020504} or state verification protocols~\cite{PhysRevApplied.12.054047,PhysRevResearch.2.043323,PhysRevLett.115.220502}. Such low-effort measurement schemes have been implemented experimentally to assess fidelities of photonic~\cite{PhysRevLett.98.180502,Lu:2007ab} and superconducting~\cite{PhysRevLett.122.110501,Mooney:2019aa} cluster states. As we show below, however, the lower bound measured in this way is not tight;  the bound deviates already to  first-order in the experimentally relevant errors, and hence does not accurately characterize the true fidelity.

In this Article, we devise a fidelity measurement scheme for cluster states, which is motivated by physical considerations and  measurable with  resources scaling only linearly with the number of qubits. Our scheme provides a lower bound for the true fidelity and thus never overestimates the quality of the generated state. As opposed to similar schemes~\cite{PhysRevLett.117.210504,PhysRevLett.94.060501,GUHNE20091,PhysRevA.76.030305,PhysRevA.72.022340,PhysRevLett.103.020504}, we exploit that errors occurring in physical systems typically have a well defined structure and affect qubits locally. As an example, cluster states can be prepared by applying two-qubit entangling gates between pairs of nearest-neighbor qubits. Hence the experimentally most likely errors are single-qubit and local two-qubit errors (errors in nearest-neighbor qubits). Targeting our measurement scheme to  such errors allows us to achieve much tighter  bounds for realistic error sources. For one-dimensional $N$-qubit cluster states, our scheme requires $\approx 11N$ measurement settings and accounts correctly for all single-qubit and local two-qubit errors to first-order in the errors. If we allow for a small inaccuracy in the first-order errors, the number of measurement settings can be reduced to $3(N-1)$. Additionally, the same measurement procedure also detects the majority of all possible higher-order errors, hence providing an excellent low-effort approximation of the true fidelity. In the case of two-dimensional cluster states, $N$ measurement settings are sufficient to account for all single-qubit errors as well as the majority of the local two-qubit errors to first-order. 

\section{Simple lower-bound fidelity}

Our approach utilizes a stabilizer description,
where the quantum state is described by a complete set of stabilizer operators. Owing to  symmetries of the state, one can then design measurement settings that simultaneously measures a large number of stabilizer operators in a single run,
dramatically reducing the amount of resources required~\cite{PhysRevLett.117.210504,PhysRevLett.94.060501,GUHNE20091,PhysRevA.76.030305,PhysRevA.72.022340,PhysRevLett.103.020504}. A one-dimensional cluster state can be defined as the simultaneous eigenstate of its stabilizer operators ${g}_i$,
\begin{equation}\label{eq:S_i}
    {g}_i\ket{\psi} = \ket{\psi} 
    \;
    \forall i=[1,N],
\end{equation}
where
\begin{equation}\label{eq:cluster_g}
    \begin{aligned}
    {g}_i 
    &= 
    {Z}_{i-1}{X}_{i}{Z}_{i+1} \; \forall i=[2,N-1]
    \\
    {g}_1 
    &= 
    {X}_{1}{Z}_{2},
    \;
    {g}_N 
    = 
    {Z}_{N-1}{X}_{N},
    \end{aligned}
\end{equation}
i.e. ${g}_i$ applies the Pauli-X operator to the $i$th qubit and Pauli-Z operators to its nearest neighbors. 

The fidelity of a density matrix ${\rho}$ is defined by its overlap         $\mathcal{F}
    =
    \mathrm{Tr}
    \Big{\{}
    \rho
    {\ket{\psi}}\bra{\psi}
    \Big{\}}$
with the ideal state $\ket{\psi}$. Using Eq.~\eqref{eq:S_i} and the fact that the product of two stabilizers is also a stabilizer, the projector to the ideal state can be expressed as
\begin{equation}\label{eq:projector_g}
    \ket{\psi}\bra{\psi}
    =
    \prod_{i=1}^{N} \frac{1 + {g}_i}{2}
    =
    \prod_{j\in \mathrm{odd}}
    {G}_j
    \prod_{i\in \mathrm{even}}
    {G}_i,
\end{equation}
where we have introduced 
${G}_i = (1+{g}_i)/2$ such that $\langle G_i \rangle = 1$ if the $i$th stabilizer is correct and $\langle G_i \rangle = 0$ otherwise. The  fidelity can therefore be determined by measuring combinations of  stabilizers ${g}_i$ rather than performing QST. While the stabilizer operators all commute and can thus in principle be measured simultaneously, most experiments rely on measuring single qubit operators. Since the stabilizers involve both $X$ and $Z$ operators, which cannot be measured simultaneously, we need to multiply out the combinations in Eq.~\eqref{eq:projector_g} and measure each combination separately. The number of such combinations scales exponentially with the system size, leading to the same intractable resource overhead as required for full tomography.

\begin{figure}[t]
\centering
\includegraphics[width=0.95\columnwidth]{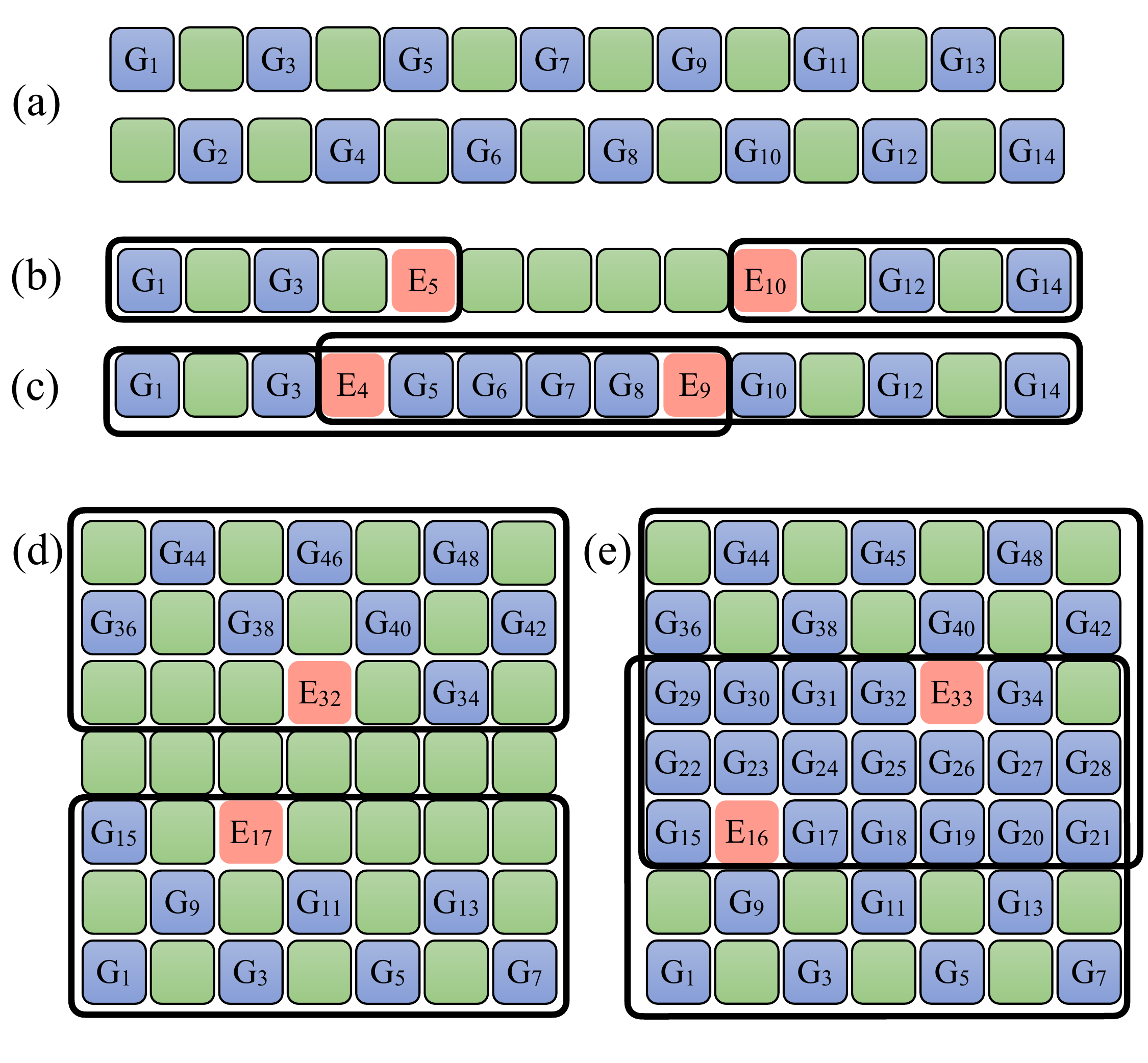}
\caption{\label{fig:1} Structure of operators required for (a)~the simple lower-bound fidelity of Eq.~\eqref{eq:proj_naive} and (b--e) the full fidelity characterization. Blue squares correspond to the operators $G_i$ centered on that site. Stabilizers  corresponding to green squares are not included. The simple lower bound~\eqref{eq:proj_naive}, which  separately measures even and odd stabilizers, thus corresponds to the two patterns in (a). This simple estimate can be improved by checking for the presence of double errors, as indicated in (b--c), where  salmon squares denote an error in that particular stabilizer. 
The two regions in black frames in (b) correspond to the different terms in Eq.~\eqref{eq:1-g}. When multiplied together as in Eq.~\eqref{eq:4th_order_expantion}, these lead to the patterns shown in  panels (b) and (c), which correspond to the first and third terms in Eq.~\eqref{eq:4th_order_expantion}, respectively, as well as other terms corresponding to the second term in Eq.~\eqref{eq:4th_order_expantion} (not shown). (b) can be measured with a single measurement setting, while (c) requires a number of measurement settings growing exponentially with the overlap region. (d,e)~Analogous to (b,c), but for a two-dimensional cluster state.}
\end{figure}

\begin{figure}[t]
\centering
\includegraphics[width=0.95\columnwidth]{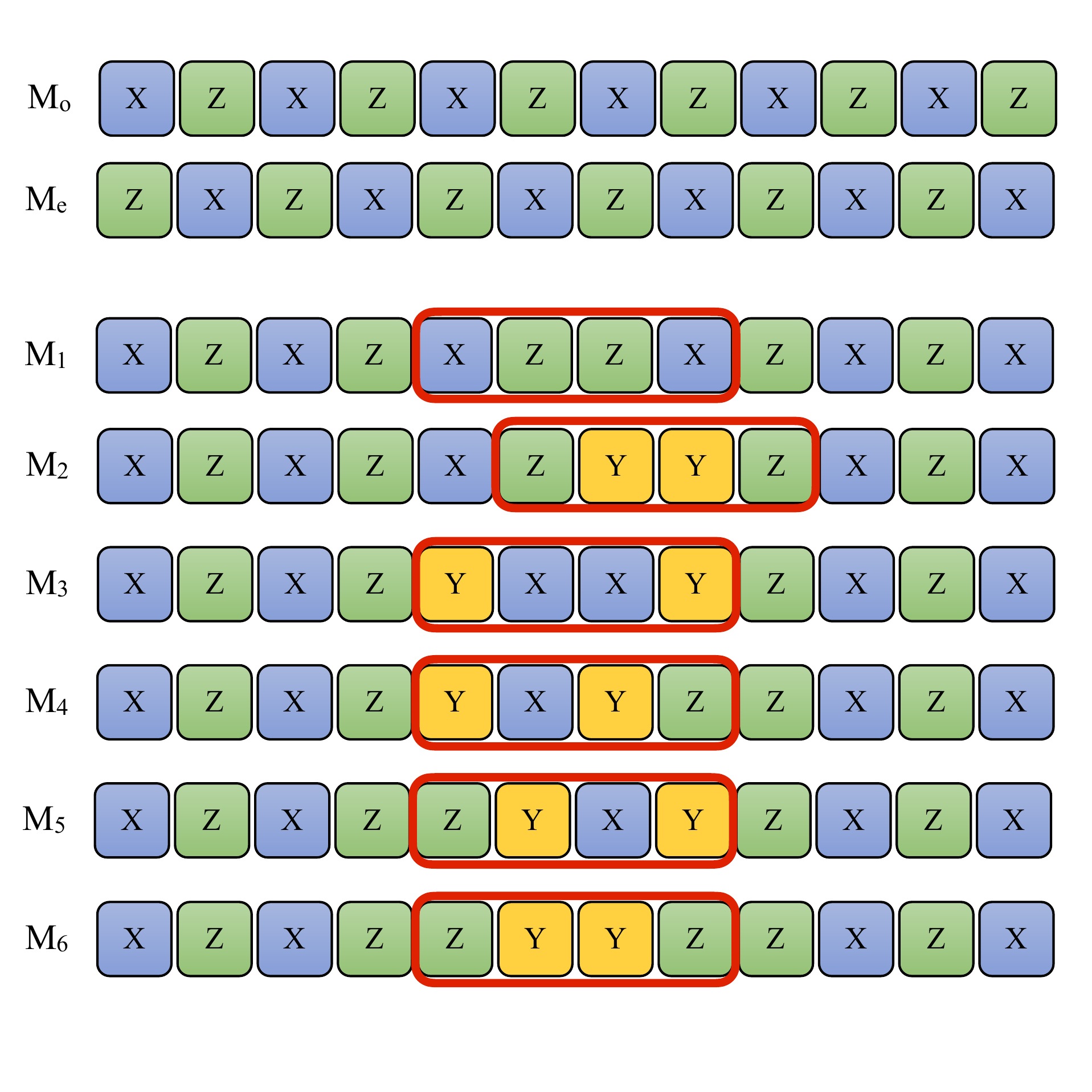}
\caption{\label{fig:2} Measurement settings used for low-effort fidelity estimations. The simple lower bound of Eq.~\eqref{eq:proj_naive} can be evaluated with two measurement settings $M_{\mathrm{o}}$ and $M_{\mathrm{e}}$. The lower bound of Eq.~\eqref{eq:refined_bound_simp}, which is tight to first order in single qubit errors and almost tight for two qubit errors, can be evaluated with the additional measurement settings $M_1$--$M_6$, with the sliding window~(red rectangle) starting at even sites for $M_2$ and odd sites for $M_1$, $M_3$--$M_6$, hence requiring $3(N-1)$ measurement settings in total.
}
\end{figure}

The resource overhead can be greatly reduced by ignoring correlations that are of little relevance for a given target state, i.e. correlations between qubits that are unlikely to appear during the state preparation. Introducing the shorthand notation $G_{\mathrm{o}}$~($G_{\mathrm{e}}$) for a product of all odd~(even) operators $G_i$, as illustrated in Fig.~\ref{fig:1}(a), the projector~\eqref{eq:projector_g} reads
\begin{equation}\label{eq:projector_g_split}
    \begin{aligned}
    \ket{\psi}\bra{\psi}
    &=
    G_{\mathrm{o}}
    +
    G_{\mathrm{e}}
    -1
    +(1-G_{\mathrm{o}})(1-G_{\mathrm{e}}).
    \end{aligned}
\end{equation}
The lower-bound fidelity of Refs.~\cite{PhysRevLett.117.210504,PhysRevLett.94.060501,GUHNE20091,PhysRevA.76.030305,PhysRevA.72.022340,PhysRevLett.103.020504} was derived by omitting the non-negative term $(1-G_{\mathrm{e}})(1-G_{\mathrm{o}})$ which yields
\begin{equation}\label{eq:proj_naive}
    \hat{P}_0
    =
    G_{\mathrm{o}} + G_{\mathrm{e}} - 1
    \leq
    \ket{\psi}\bra{\psi}.
\end{equation}
Since $G_{\mathrm{e}}$ ($G_{\mathrm{o}}$) only involves $X$ ($Z$) operators on even sites and $Z$ ($X$) operators on odd sites it can be measured with a single measurement setting. Therefore, the operator $\hat{P}_0$ can be measured using only two settings $M_{\mathrm{o}}$ and $M_{\mathrm{e}}$ illustrated in Fig.~\ref{fig:2}, providing a means for low-effort fidelity estimation. 

To investigate the performance of various  fidelity bounds, we will consider  simplified error models where we randomly apply Pauli operators with a certain probability. Although real noise sources may be more complicated, we show in Appendix~\ref{sec:arbitrary_errors} that any single and two-qubit error models can be accurately mapped to this setting to lowest order. For the lower bound~\eqref{eq:proj_naive} 
we now assume  that a Pauli error occurs in a single qubit with a probability $p$, hence introducing a state infidelity $\mathcal{I} = 1 - \mathcal{F} = p$. Under Pauli-$X$ or -$Z$ errors, the expectation value of the lower-bound operator~\eqref{eq:proj_naive} returns the correct infidelity $\mathcal{I}_{\mathrm{b}} = p$ since either the even or odd part is affected by such errors. Pauli-$Y$ errors, on the other hand, turn both observables $\langle G_{\mathrm{o}} \rangle$ and $\langle G_{\mathrm{e}} \rangle$ in Eq.~\eqref{eq:proj_naive} into zero with a probability $p$, indicating an infidelity $\mathcal{I}_{\mathrm{b}} = 2p$. Therefore, the simple lower bound of Eq.~\eqref{eq:proj_naive} is not tight to first-order in $p$ even for single-qubit errors. Analogously, any multiqubit error flipping both even and odd stabilizers results in twice the true infidelity when measured according to Eq.~\eqref{eq:proj_naive}. We note that a similar argument applies to any fidelity certification protocol that uses only two measurement settings, such as the protocols of Refs.~\cite{PhysRevResearch.2.043323,PhysRevLett.115.220502}. A single-qubit Pauli-$Y$ error occurring with probability $p$ will result in the same measurement patters as single-qubit Pauli-$X$ and $Z$ errors, each occurring with probability $p$. No classical post-processing is able to distinguish between the two cases, while the true infidelity differs by a factor of two. Hence, any protocol for determining the state fidelity based on two measurement settings $M_{\textrm{e}}$ and $M_{\textrm{o}}$ of Fig.~\ref{fig:2} will not be tight even to first order in $p$. 

\section{Refined fidelity measure}

 Our aim is to construct a physically motivated  lower-bound fidelity that correctly accounts for experimentally relevant errors to first order by measuring additional correlations while keeping the number of measurement settings linear in the system size.
 The term $(1-G_{\mathrm{o}})(1-G_{\mathrm{e}})$ in Eq.~\eqref{eq:projector_g_split} corrects the error overestimation discussed above and hence we need a procedure to lower bound it.
 
 First we notice that the term $1-G_{\mathrm{o}}$~(or $1-G_{\mathrm{e}}$) is non-zero only when errors are present  in one or more odd~(even) stabilizers. Introducing a new operator $E_i = 1-G_i$ that corresponds to an error in the $i$th stabilizer, we can express all possible errors in even and odd stabilizers as
\begin{equation}\label{eq:1-g}
    \begin{aligned}
        1-G_{\mathrm{e}}
        &=
        \sum_{i \in \mathrm{even}}
        E_i
        \prod_{\substack{k \in \mathrm{even} \\ k>i}}
        G_{k}
        \\
        1-G_{\mathrm{o}}
        &=
        \sum_{i \in \mathrm{odd}}
        E_i
        \prod_{\substack{k \in \mathrm{odd} \\ k<i}}
        G_{k},
     \end{aligned}
\end{equation}
respectively. To prove the first equation in~\eqref{eq:1-g}, we express $G_{\mathrm{e}}$ as
\begin{equation}
    \begin{aligned}
    G_{\mathrm{e}} 
    &=
    \prod_{i \in \mathrm{even}}
    G_i
    = \prod_{i \in \mathrm{even}} (1-E_i)
    \\&=
    1 - 
    \sum_{i \in \mathrm{even}}
    E_i
    +
    \sum_{\substack{i,k\in \mathrm{even} \\ k>i}}
    E_i
    E_k
    -...,
    \end{aligned}
\end{equation}
and consequently
\begin{equation}\label{eq:form_1}
\begin{aligned}
    1-G_{\mathrm{e}}
    &=
    \sum_{i \in \mathrm{even}}
    E_i
    -
    \sum_{\substack{i,k\in \mathrm{even} \\ k>i}}
    E_i
    E_k
    +
    ...
\end{aligned}
\end{equation}
Next we notice that the expression above is identical to 
\begin{equation}\label{eq:form_2}
\begin{aligned}
    &\sum_{i \in \mathrm{even}}
    E_i
    \prod_{
    \substack{j\in \mathrm{even} \\ k>i}
    }
    G_{k}
    =
    \sum_{i \in \mathrm{even}}
    E_i
    \prod_{\substack{k\in \mathrm{even} \\ k>i}}
    (1-E_k)
    \\&=
    \sum_{i \in \mathrm{even}}
    E_i
    -
    \sum_{\substack{i,k\in \mathrm{even} \\ k>i}}
    E_i
    E_k
    +
    ...
\end{aligned}
\end{equation}
Comparing these expressions we find the first line in Eq. \eqref{eq:1-g}. 

Since the operators $G_i$ and $E_i = 1-G_i$ correspond to the correct and erroneous $i$th stabilizer, respectively, the above result can be understood as grouping terms in $G_{\mathrm{e}}$ according to the rightmost error. The left hand side contain all possible errors in $G_{\mathrm{e}}$, but the right hand side groups these terms according to the rightmost error $E_i$ by demanding that there are no errors to the right of position $i$ so that $G_k=1$ for $k>i$. A single term in the sum above is illustrated by the rightmost black rectangle in Fig.~\ref{fig:1}(b) and corresponds to an erroneous $i$th stabilizer followed by error-free even stabilizers. 

Applying an analogous procedure to the term $1-G_{\mathrm{o}}$ yields the second line in Eq.~\eqref{eq:1-g}. As opposed to the first line,  we here count the errors starting from the opposite end, i.e. we express $1-G_{\mathrm{o}}$ as a sum of erroneous stabilizers preceded by error-free odd stabilizers. A term in the sum above corresponds to the leftmost black rectangle in Fig.~\ref{fig:1}(b). In short, Eqs.~\eqref{eq:1-g} divide all possible errors for odd (even) stabilizers into groups according to the position of the first (last) error. 

With expressions~\eqref{eq:1-g}, the last term of Eq.~\eqref{eq:projector_g_split} yields
\begin{equation} \label{eq:4th_order_expantion}
    \begin{aligned}
    &(1-G_{\mathrm{o}})(1-G_{\mathrm{e}}) 
    \\&= 
    \sum_{i \in \mathrm{odd}}
    \Big{[}
    \sum_{\substack{j \in \mathrm{even} \\ j-i >  3}}
    E^{(2)}_{ij}
    +
    \sum_{\substack{j \in \mathrm{even} \\ |i-j| \leq 3}}
    E^{(2)}_{ij}
    +
    \sum_{\substack{j \in \mathrm{even} \\ i-j > 3}}
    E^{(2)}_{ij}
    \Big{]},
    \end{aligned}
\end{equation}
where
\begin{equation}\label{eq:E_ij}
    E^{(2)}_{ij}
    =
    E_i
    E_j
    \prod_{\substack{k\in \mathrm{odd} \\ k<i}}
    G_{k}
    \prod_{\substack{m \in \mathrm{even} \\ m>j}}
    G_{m}.
\end{equation}
Each of the $\sim N^2$ terms in the sums represents a positive contribution to the fidelity. Measuring a term and adding it to the simple bound in Eq.~\eqref{eq:proj_naive} thus always brings us closer to the true fidelity, which we approach from below. This allows us to design measurement setting tailored to  measure those terms, which we consider to be  most important based on physical grounds.  

\begin{figure}[t]
\centering
\includegraphics[width=0.95\columnwidth]{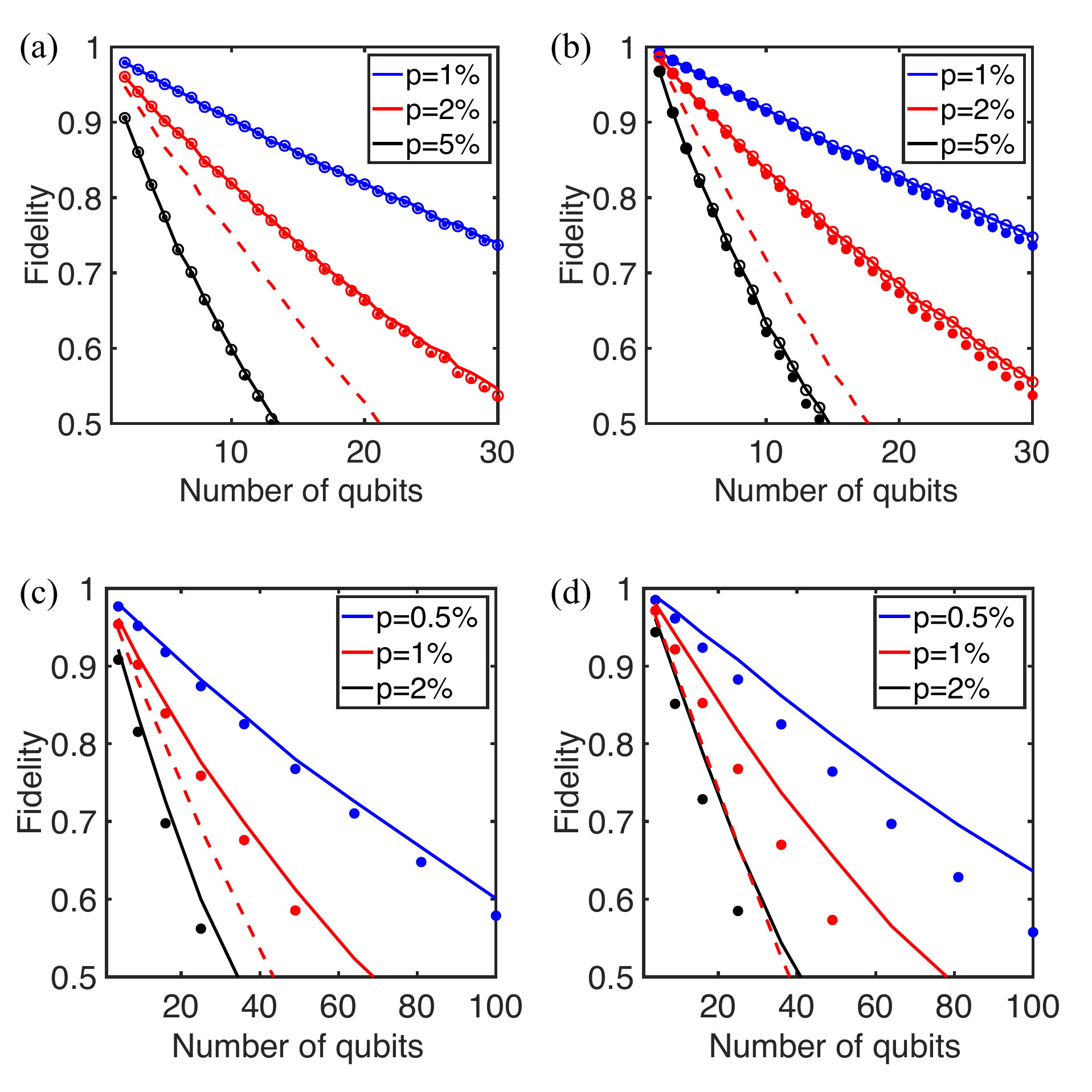}
\caption{\label{fig:3} 
Comparison between the full fidelity~(solid lines) and the lower-bound fidelities versus the number of qubits for various error probabilities $p$. Panel (a)~[(b)] corresponds to one-dimensional states affected  by single-qubit~[local two-qubit] errors applied to each qubit~[nearest-neighbor pair of qubits] with a probability $p$. Circles~[dots] correspond to the lower-bound fidelity of Eq.~\eqref{eq:refined_bound}~[\eqref{eq:refined_bound_simp}] that require $\approx 11N$~[$3(N-1)$] measurement settings. 
For comparison, we also show the simple lower bound \eqref{eq:proj_naive} requiring only two measurement settings for $p=2$\%~(dashed lines). (c,d)~Analogous to (a,b), but for a two-dimensional square cluster state and the lower-bound fidelity of Eq.~\eqref{eq:2D_refined}~(dots).} 
\end{figure}

In Eq.~\eqref{eq:4th_order_expantion} we have divided all possible errors that simultaneously flip even and odd stabilizers into three terms. The first term contains products of non-overlapping even and odd operators, such as the one shown in Fig.~\ref{fig:1}(b). With the measurement setting $M_1$ depicted in Fig.~\ref{fig:2}, all such  operators with $i$ to the left and $j$ to the right of the red rectangle can be measured. By sliding the rectangle we get all such terms and the total number of measurement settings thus scales linearly with $N$, as shown in Appendix~\ref{sec:1D_settings}.

The second term contains products of two flipped stabilizers $E_i$ and $E_j$ located in a close proximity to each other, which can be realized by either a single Pauli-$Y$ error or errors in a pair of nearest-neighbor qubits. Therefore, this term describes the experimentally most relevant errors, as discussed in the introduction. Since this term only involve a limited number of nearby stabilizers, we can multiply these out and get a set of operators, which can be measured by a limited number of measurement settings. As detailed in Appendix~\ref{sec:1D_settings} this can be achieved with the measurement settings $M_1-M_6$ depicted in Fig.~\ref{fig:2} as well as some similar terms. The total number of such measurement settings again scales linearly in $N$.

Finally, the last sum in Eq.~\eqref{eq:4th_order_expantion} corresponds to a large overlap between different terms of the operators $1-G_{\mathrm{o}}$ and $1-G_{\mathrm{e}}$, similar to the one in Fig.~\ref{fig:1}(c). 
This term leads to a number of measurement settings scaling exponentially with the length of the overlap region since it involves products of many neighboring stabilizers with different Pauli operators. On the other hand, it corresponds to a small subset of the non-local multiqubit errors and is unlikely to occur experimentally since the errors are far apart. Hence, by omitting this term one can reduce the number of measurement settings to linear in $N$ while introducing only a small  error. Ignoring this term and substituting Eq.~\eqref{eq:4th_order_expantion} into Eq.~\eqref{eq:projector_g_split} yields the refined lower bound
\begin{equation}\label{eq:refined_bound}
    \begin{aligned}
    \hat{P}_{1D}
    &=
    G_{\mathrm{e}}
    +
    G_{\mathrm{o}}
    -1
    +
    \sum_{\substack{i \in \mathrm{odd} \\ j \in \mathrm{even} \\ j \geq i-3}}
    E_i
    E_j
    \prod_{\substack{k\in \mathrm{odd} \\ k<i}}
    G_{k}
    \prod_{\substack{m \in \mathrm{even} \\ m>j}}
    G_{m}.
    \end{aligned}
\end{equation}
To first order in local errors, this operator is identical to the full projector $\ket{\psi}\bra{\psi}$ for the states affected by  single-qubit or  local two-qubit errors and can be evaluated using $\approx 11N$ measurement settings provided in Appendix~\ref{sec:1D_settings}. 

The number of measurement settings can be further reduced by omitting the term with $j=i-3$ in the equation above, which yields the simplified lower bound
\begin{equation}\label{eq:refined_bound_simp}
    \begin{aligned}
    \hat{P}'_{1D}
    &=
    G_{\mathrm{e}}
    +
    G_{\mathrm{o}}
    -1
    +
    \sum_{\substack{i \in \mathrm{odd} \\ j \in \mathrm{even} \\ j > i-3}}
    E_i
    E_j
    \prod_{\substack{k\in \mathrm{odd} \\ k<i}}
    G_{k}
    \prod_{\substack{m \in \mathrm{even} \\ m>j}}
    G_{m}.
    \end{aligned}
\end{equation}
As we show in Appendix~\ref{sec:1D_settings}, this simplification greatly reduces the number of measurement settings to the $3(N-1)$ settings shown in Fig~\ref{fig:2}. At the same time, it only introduces a small deviation from the true fidelity. Assuming that errors are described by the two-qubit depolarizing channel with equally probable single- and local two-qubit errors, Eq.~\eqref{eq:refined_bound_simp} catches $\approx 96\%$ of the first order error measured by Eq.~\eqref{eq:refined_bound}. 

To study the performance of the lower bound~\eqref{eq:refined_bound_simp} beyond the first-order approximation, we consider the single qubit depolarizing channel, where one of three Pauli errors is  applied to each qubit with an equal probability $p$. The full fidelity and the simplified lower bound of Eq.~\eqref{eq:refined_bound_simp} are calculated using Monte Carlo simulations and shown in Fig.~\ref{fig:3}(a). It agrees well with the analytical solutions provided in Appendix~\ref{sec:analytical}. Figure.~\ref{fig:3}(b) shows an analogous plot for a state affected by equally probable  local two-qubit errors, such that $P(\sigma^{i}_l \sigma^{i+1}_m) = p/9$ with $l,m \in \{ x,y,z \}$. Evidently, the proposed fidelity measure is an excellent approximation of the true fidelity for these realistic noise sources. 

\section{Two-dimensional cluster states}

Our measurement scheme can be directly generalized to two-dimensional  cluster states, which  are of tremendous interest  as resources for measurement-based quantum computation~\cite{RevModPhys.79.135,PhysRevLett.93.040503,PhysRevLett.95.010501,PhysRevLett.95.010501,Knill:2001aa,Azuma:2015aa,Li:2019aa,PhysRevX.10.021071}.  Labelling qubits with a single index as in Fig.~\ref{fig:1}(d,e), the correlation term $(1-G_{\mathrm{o}})(1-G_{\mathrm{e}})$ takes a form similar to Eq.~\eqref{eq:4th_order_expantion},
\begin{equation} \label{eq:4th_order_expantion_2D}
    \begin{aligned}
    (1-G_{\mathrm{o}})(1-G_{\mathrm{e}}) 
    &= 
    \sum_{\substack{i \in \mathrm{odd} \\ j \in \mathrm{even}}}
    E_i
    E_j
    \prod_{\substack{k\in \mathrm{odd} \\ k<i}}
    G_{k}
    \prod_{\substack{m \in \mathrm{even} \\ m>j}}
    G_{m}
    \\&=
    \sum_{i \in \mathrm{odd}}
    \Big{[}
    \sum_{\substack{j \in \mathrm{even} \\ j-i \geq N_x}}
    {E}^{(2)}_{ij}
    +
    \sum_{\substack{j \in \mathrm{even} \\ j-i < N_x}}
    {E}^{(2)}_{ij}
    \Big{]},
    \end{aligned}
\end{equation}
where $N_x$ is the horizontal size of the state. In the second line of Eq.~\eqref{eq:4th_order_expantion_2D} we have divided all errors into two categories. The first sum corresponds to the terms in $(1-G_{\mathrm{o}})$ and $(1-G_{\mathrm{e}})$ that are at least one layer apart from each other, e.g. as in Fig.~\ref{fig:1}(d). Consequently, the second sum includes the terms that are less than $N_x$ closer to each other or overlap, such as the one shown in Fig.~\ref{fig:1}(e). This second term requires a number of measurement settings that grows approximately exponentially with at least one dimension of the cluster, i.e. $\propto \mathrm{exp}(\sqrt{N})$ for a square cluster, see Appendix~\ref{sec:2D_states} for detailed discussion. We shall therefore ignore this term. 

Substituting Eq.~\eqref{eq:4th_order_expantion_2D} into Eq.~\eqref{eq:projector_g_split} and omitting the last term yields a projector
\begin{equation}
\begin{aligned}
\label{eq:2D_refined}
    &\hat{P}_{2D}
    =
    G_{\mathrm{e}}
    +
    G_{\mathrm{o}}
    -1
    \\&+
    \sum_{i \in \mathrm{odd}}
    \sum_{\substack{j \in \mathrm{even} \\ j-i \geq N_x}}
    E_i
    E_{j}
    \prod_{\substack{k\in \mathrm{odd} \\ k<i}}
    G_{k}
    \prod_{\substack{m \in \mathrm{even} \\ m>j}}
    G_{m},
\end{aligned}
\end{equation}
which can be measured with $\sim N$ settings as we discuss in detail in Appendix~\ref{sec:2D_states}. As with the one-dimensional case of Eq.~\eqref{eq:refined_bound_simp}, the equation above yields the exact fidelity for single-qubit errors to first-order in the error and shows excellent performance beyond that approximation, see Fig.~\ref{fig:3}(c). Furthermore, the refined lower bound~\eqref{eq:2D_refined} correctly takes into account the majority, but not all of the possible local two-qubit errors to first order. As we show in Appendix~\ref{sec:2D_states}, with equiprobable Pauli errors in adjacent qubits, the lower-bound fidelity~\eqref{eq:2D_refined} correctly detects $\approx85$\% of local two-qubit errors, which agrees well with Fig.~\ref{fig:3}(d) obtained from Monte Carlo simulations. Note that the model of equally probable two-qubit errors considered here is the worst possible scenario. Having some prior knowledge of the error sources, the measurement scheme can be optimized to identify the most likely errors. 

\section{Conclusion}

In conclusion, we have developed a scheme for determination of the cluster-state fidelity with a number of measurement settings scaling linearly in the system size. The constructed fidelity measure is a strict lower bound of the true fidelity and provides an excellent approximation for the true fidelity of one-dimensional cluster states affected by realistic noise sources. The scheme also performs well for  states of higher dimension, showing only a small deviation when the state is affected by certain multiqubit errors. 
The method proposed in this Article originates from a simple idea: we group the errors in a convenient way, such that the hardest detectable correlations are also the least probable ones and can therefore be ignored. This idea can be used as an inspiration for constructing low-effort fidelity measures~(or other observables) for different classes of quantum states. Owing to its simplicity, high accuracy, and flexibility, we expect this method to play an important role in verification of multiqubit states  in near-future experiments. 

\begin{acknowledgments}

We gratefully acknowledge financial support from Danmarks Grundforskningsfond~(DNRF 139, Hy-Q Center for Hybrid Quantum Networks) and the European Union Horizon 2020 research and innovation programme under grant agreement N\textsuperscript{\underline{o}}~820445 and project name Quantum Internet Alliance. The authors thank Peter Lodahl, Martin H. Appel, and Alexey Tiranov for inspiring discussions.
\end{acknowledgments}

\bibliographystyle{apsrev4-1}
\bibliography{reflist}

\clearpage
\onecolumngrid
\appendix

\section{Arbitrary errors}\label{sec:arbitrary_errors}

In the main text we consider an $N$-qubit cluster state, and a noise model where qubits are affected by either Pauli $X$, $Y$, or $Z$ errors. Below we show that the analysis provided in the main text immediately generalizes to  arbitrary noise channels for the dominant single error term, which is our main interest here. Hence the description is also applicable when qubits are affected by an arbitrary superposition of $X$, $Y$, and $Z$ errors. In particular, we show that a general noise channel turns into a statistical mixture of orthogonal noise channels when acting on a cluster state. 

Under a general noise channel the density matrix $\rho_0 = \ket{\Psi}\bra{\Psi}$ transforms according to
\begin{equation}
\label{eq:noise}
    \rho 
    =
    \sum_l
    E_l
    \rho_0
    E^{\dagger}_l,
\end{equation}
where
$E_l$ are Kraus operators describing different noise processes. Since we are mainly interested in a single error occurring in the system and  the channel consists of   a statistical mixture of different Kraus operators, is it sufficient to consider only a single Kraus operator $E$ without loss of generality. For a single qubit error we write this operator acting on a chosen qubit as a superposition of Pauli errors,
\begin{equation}
\label{eq:E}
    E
    =
    \alpha
    +
    \sum_{j=x,y,z}
    \beta_j
    \sigma_j,
\end{equation}
where we omit the qubit index for brevity. The probability of such a channel is
\begin{equation}
    P
    =
    \bra{\Psi}
    E^{\dagger}
    E
    \ket{\Psi}
    =
    |\alpha|^2
    +
    \sum_{j=x,y,z}
    |\beta_j|^2,
\end{equation}
where we used the fact that $\ket{\Psi}$ is a stabilizer state, and that any operator which is not in its stabilizer group takes $\ket{\Psi}$ to an orthogonal state. Products of Pauli operators $\sigma_j\sigma_k$ yield another Pauli operator, which is not in the stabilizer group unless $j=k$. Hence all terms with $j\neq k$ vanish. Since the right hand side of this expression can be interpreted as a sum of probabilities, an arbitrary superposition of Pauli errors therefore becomes a statistical mixture of Pauli errors $\sigma_j$ occurring with probability $|\beta_j|^2$ when acting on a stabilizer state such as a cluster state. Furthermore, because any single-qubit error takes a cluster state to an orthogonal state, the exact fidelity reads
\begin{equation}
    \mathcal{F}
    =
    |\bra{\Psi} E \ket{\Psi}|^2
    =
    |\alpha|^2.
\end{equation}
Hence the fidelity in this case is just equal to the probability that no error occur. 

For all  the lower-bound fidelities considered, we need to calculate expectation values of  operators of the form $\prod_{i \in I}g_i$, where $I$ is some subset of the stabilizers generators. Applying similar arguments as above  we get 
\begin{equation}\label{eq:single_qubit}
    \begin{aligned}
    &\mathrm{Tr}
    \Big{\{}
    \prod_{i\in I}
    g_i
    \rho
    \Big{\}}
    =
    \bra{\Psi}
    E^{\dagger}
    \prod_{i\in I}
    g_i
    E
    \ket{\Psi}
    =
    \bra{\Psi}
    \Big{(}
    \alpha^*
    +
    \sum_{j=x,y,z}
    \beta_j^*
    \sigma_j
    \Big{)}
    \prod_{i\in I}
    g_i
    \Big{(}
    \alpha
    +
    \sum_{k=x,y,z}
    \beta_k
    \sigma_k
    \Big{)}
    \ket{\Psi}
    \\&=
    |\alpha|^2
    +
    \alpha^*
    \sum_{j=x,y,z}
    \beta_j
    \underbrace{
    \bra{\Psi}
    \sigma_j
    \ket{\Psi}}_{=0}
    +
    \alpha
    \sum_{k=x,y,z}
    \beta_k^*
    \underbrace{
    \bra{\Psi}
    \sigma_k
    \ket{\Psi}}_{=0}
    +
    \sum_{j,k = x,y,z}
    \beta_j
    \beta_k^*
    \bra{\Psi}
    \sigma_j
    \prod_{i\in I}
    g_i
    \sigma_k
    \ket{\Psi}
    \\&
    =
    |\alpha|^2
    +
    \sum_{j=x,y,z}
    |\beta_j|^2
    \mathrm{Tr}
    \Big{\{}
    \sigma_j
    \ket{\Psi}
    \bra{\Psi}
    \sigma_j
    \prod_{i\in I}
    g_i
    \Big{\}}.
    \end{aligned}
\end{equation}
The central step in this argument is  the reduction of  the last term on the second line to the final term in the third line. Here we have used that since $\prod_i g_i$ can be written as a product of Pauli-operators, $\sigma_j$ will either commute or anti-commute with $\prod_i g_i$. Hence we have $\sigma_j\prod_i g_i\sigma_k=
\pm \prod_i g_i\sigma_j\sigma_k$. Since $|\Psi\rangle$ is an eigenstate of all  $g_i$-operators with eigenvalue $+1$, we can now remove the term $\prod_i g_i$ from the expectation value. If $j\neq k$ the combination  $\sigma_j\sigma_k$ is another Pauli-operator, which is not in the stabilizer group. The remaining expression $\pm\langle\Psi|\sigma_j\sigma_k|\Psi\rangle$ thus  vanish since a non-stabilizer operator
$\sigma_j\sigma_k$  take $\ket{\Psi}$ to an orthogonal state $\ket{\Psi'}$. Therefore, only terms with $j=k$ survive in the last line of the equation, which is equivalent to a statistical mixture of Pauli-errors.

The discussion above easily generalizes to any two-qubit nearest-neighbor error. The general two-qubit error between qubits $m$ and $n$ reads
\begin{equation}\label{eq:2_qubit_general}
    E
    =
    \sum_{j,j' = 0,x,y,z}
    \beta_{jj'}
    \sigma_j^m
    \sigma_{j'}^{n}
\end{equation}
where $\sigma_0$ is the identity matrix. Using similar arguments as above,  error terms $\sigma_j^m
    \sigma_{j'}^{n} \prod_i g_i\sigma_k^m
    \sigma_{k'}^{n}$
can be (anti)commuted to yield products of two Pauli operators,
\begin{equation}
    \begin{aligned}
    &\sum_{j,j'=0,x,y,z}
    \sum_{k,k'=0,x,y,z}
    \beta_{jj'}
    \beta_{kk'}^*
    \bra{\Psi}
    \sigma_j^m
    \sigma_{j'}^{n}
    \prod_i g_i
    \sigma_k^m
    \sigma_{k'}^{n}
    \ket{\Psi}
    \\&=
    \pm
    \sum_{j,j'=0,x,y,z}
    \sum_{k,k'=0,x,y,z}
    \beta_{jj'}
    \beta_{kk'}^*
    \bra{\Psi}
    \sigma_j^m
    \sigma_{j'}^{n}
    \sigma_k^m
    \sigma_{k'}^{n}
    \ket{\Psi}
    \\&=
    \pm
    \sum_{j,j'=0,x,y,z}
    \sum_{k,k'=0,x,y,z}
    \beta_{jj'}
    \beta_{kk'}^*
    e_{jks}
    e_{j'k's'}
    \bra{\Psi}
    \sigma_s^m
    \sigma_{s'}^{n}
    \ket{\Psi},
    \end{aligned}
\end{equation}
where $e_{jks}$ is the Levi-Civita symbol. Since cluster states in 1D (2D) are stabilized by operators, which are product of three~(five) Pauli operators acting on neighbouring qubits, all terms in the sum above vanish unless $\sigma_s^m = \sigma_0$ and $\sigma_{s'}^n = \sigma_0$, i.e. unless $j=k$ and $j'=k'$. Therefore, we arrive at 
\begin{equation}
    \begin{aligned}
    &\mathrm{Tr}
    \Big{\{}
    \prod_{i\in I}
    g_i
    \rho
    \Big{\}}
    =
    \sum_{j=0,x,y,z}
    \sum_{j'=0,x,y,z}
    |\beta_{jj'}|^2
    \mathrm{Tr}
    \Big{\{}
    \sigma_j^{m}
    \sigma_{j'}^{n}
    \rho
    \sigma_j^{m}
    \sigma_{j'}^{n}
    \prod_{i\in I}
    g_i
    \Big{\}}.
    \end{aligned}
\end{equation}
Again, the general two-qubit error becomes a statistical mixture of orthogonal errors, and hence the analysis of our measurement scheme is valid for a general noise of the form~\eqref{eq:noise}. The only exception to the above argument is two particular errors at the edges of   cluster states in 1D. Here we have the stabilizers $X_1Z_2$ and $Z_{N-1}X_N$ which are only two particle operators. This gives a small correction $\propto 1/N$. Any other local two-qubit error takes a cluster state to an orthogonal state. 

\section{Number of measurement settings, 1D cluster state}\label{sec:1D_settings}

\begin{figure}[h!]
\centering
\includegraphics[width=0.65\columnwidth]{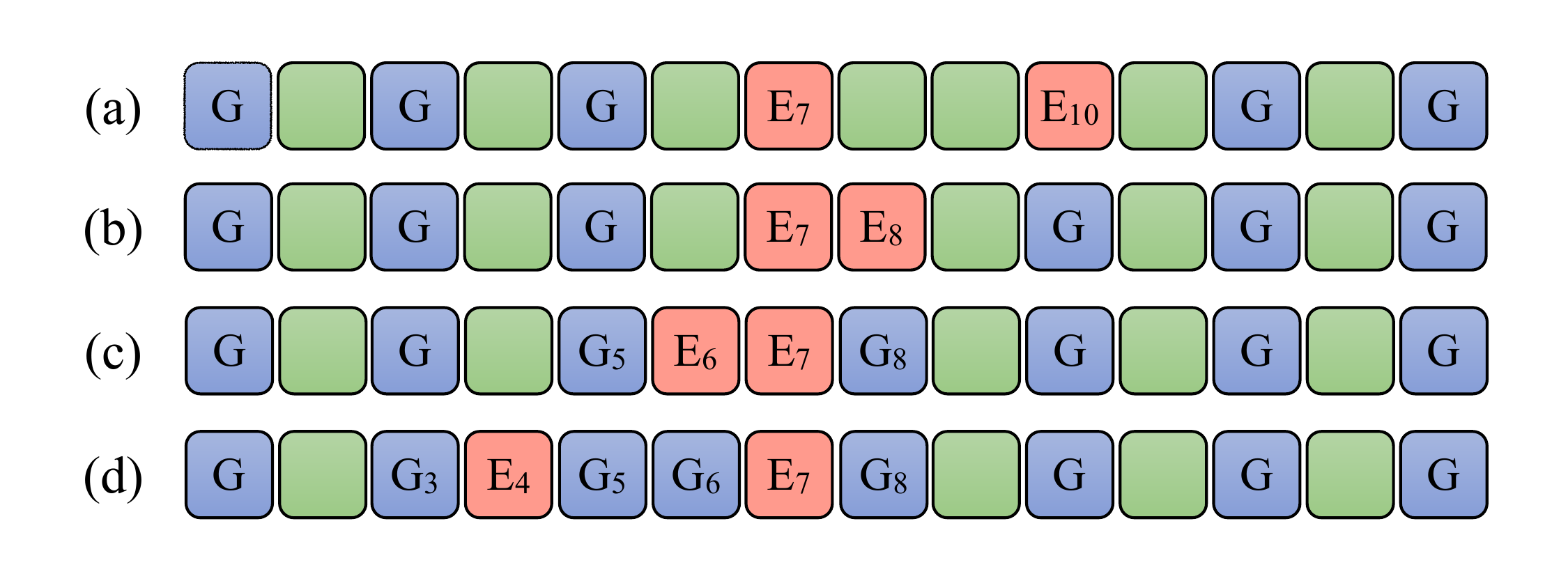}
\caption{\label{fig:SI1} Schematic representation of the corresponding terms of Eq.~\eqref{eq:si:4th_order} for $i=7$.} 
\end{figure}

Here we evaluate the number of measurement settings required for applying our procedure to 1D cluster states. To measure the refined lower bound of Eq.~\eqref{eq:refined_bound}, one needs to measure the expectation value of the operators
\begin{equation}\label{eq:si:4th_order}
    \begin{aligned}
    \sum_{\substack{i \in \mathrm{odd} \\ j \in \mathrm{even} \\ j \geq i-3}}
    E_i
    E_j
    \prod_{\substack{k\in \mathrm{odd} \\ k<i}}
    G_{k}
    \prod_{\substack{m \in \mathrm{even} \\ m>j}}
    G_{m}
    &=
    \underbrace{\sum_{\substack{i \in \mathrm{odd} \\ j \in \mathrm{even} \\ j>i+1}}
    E_i
    E_j
    \prod_{\substack{k\in \mathrm{odd} \\ k<i}}
    G_{k}
    \prod_{\substack{m \in \mathrm{even} \\ m>j}}
    G_{m}}_{(a)}
    +
    \underbrace{\sum_{i \in \mathrm{odd}}
    E_i
    E_{i+1}
    \prod_{\substack{k\in \mathrm{odd} \\ k<i}}
    G_{k}
    \prod_{\substack{m \in \mathrm{even} \\ m>i+1}}
    G_{m}}_{(b)}
    \\&+
    \underbrace{\sum_{i \in \mathrm{odd}}
    E_i
    E_{i-1}
    \prod_{\substack{k\in \mathrm{odd} \\ k<i}}
    G_{k}
    \prod_{\substack{m \in \mathrm{even} \\ m>i-1}}
    G_{m}}_{(c)}
    +
    \underbrace{\sum_{i \in \mathrm{odd}}
    E_i
    E_{i-3}
    \prod_{\substack{k\in \mathrm{odd} \\ k<i}}
    G_{k}
    \prod_{\substack{m \in \mathrm{even} \\ m>i-3}}
    G_{m}}_{(d)}.
    \end{aligned}
\end{equation}
The terms (a)--(d) above are shown schematically in the corresponding panels of Fig.~\ref{fig:SI1} for $i=7$ [for simplicity we only include the $j=10$ term in (a)].  By omitting the term (d), we arrive at the simplified lower bound of Eq.~\eqref{eq:refined_bound_simp}. As we will see below, this simplification reduces the number of measurement settings by a factor of $\sim 11/3$, while introducing only a small inaccuracy for realistic experimental errors. 

\begin{figure}[h!]
\centering
\includegraphics[width=0.6\columnwidth]{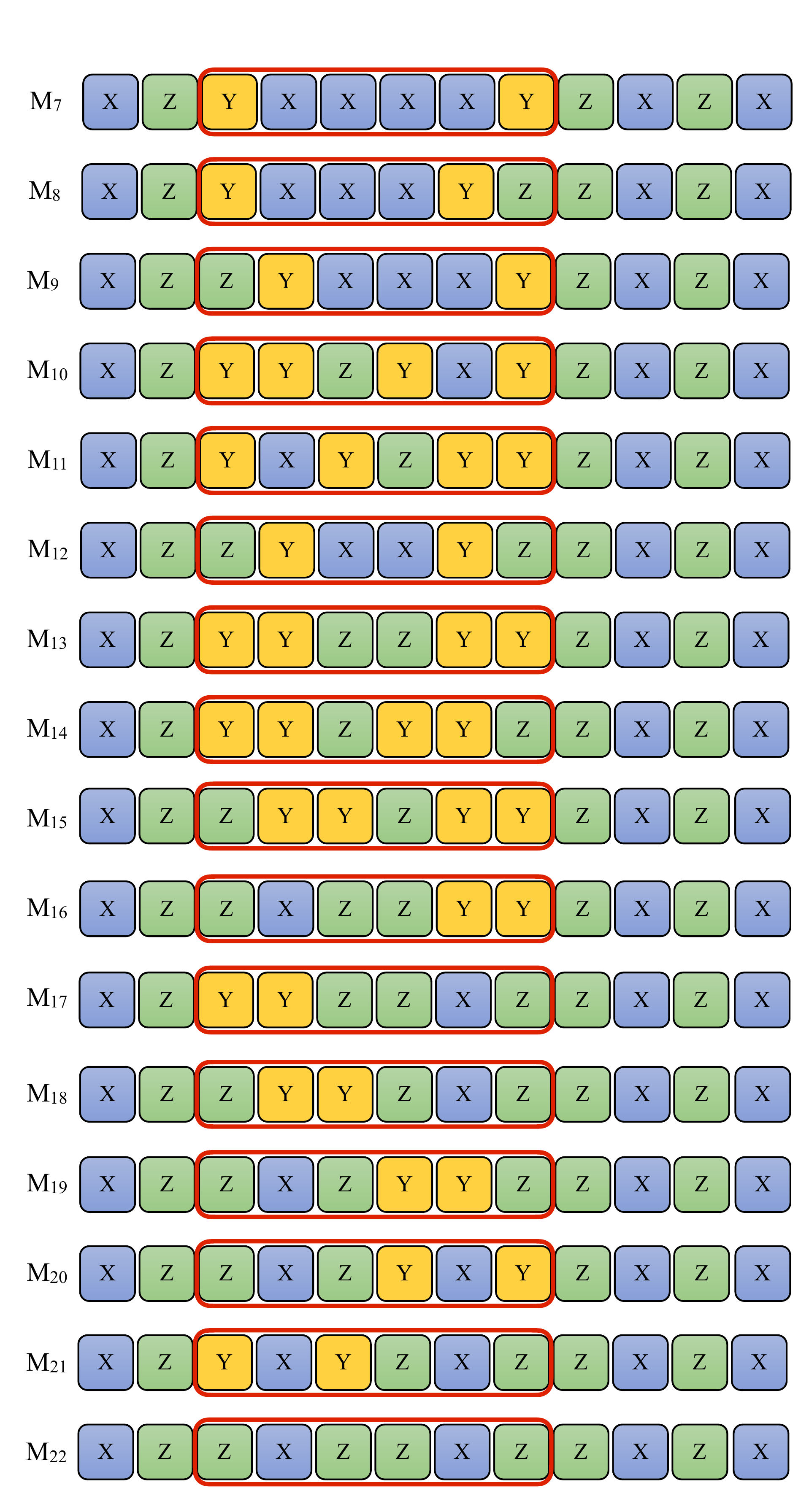}
\caption{\label{fig:SI9} Measurement settings $M_7-M_{22}$ required for determining the term (d) in Eq.~\eqref{eq:si:4th_order}, i.e. the first-order fidelity of Eq.~\eqref{eq:refined_bound}. The red rectangles slide with a step 2, hence resulting in $8N$~(for large $N$) measurement settings.} 
\end{figure}

An operator corresponding to panel (a) of Fig.~\ref{fig:SI1} can be measured with a setting of the type $M_1$ shown in Fig.~\ref{fig:2}. Here we can simultaneously measure all even stabilisers to the right and all odd stabilizers to the left of the two repeating $Z$ operators.

To measure term (b) of Eq.~\eqref{eq:si:4th_order} we expand the operator of Fig.~\ref{fig:SI1}(b) as
\begin{equation}
    \Big{(}
    \prod_{\mathrm{left}}G
    \Big{)}
    E_7
    E_8
    \Big{(}
    \prod_{\mathrm{right}}G
    \Big{)}
    \propto
    \underbrace{
    \Big{(}
    \prod_{\mathrm{left}}G
    \Big{)}
    \Big{(}1 - g_7 - g_8\Big{)}
    \Big{(}
    \prod_{\mathrm{right}}G
    \Big{)}
    }_{M_1}
    +
    \underbrace{
    \Big{(}
    \prod_{\mathrm{left}}G
    \Big{)}
    g_7g_8
    \Big{(}
    \prod_{\mathrm{right}}G
    \Big{)}}_{M_2},
\end{equation}
where left~(right) denote odd~(even) indices preceding~(following) the operator $E_7E_8$. The first term on the RHS of this equation can be measured using two settings of the type $M_1$ shown in Fig.~\ref{fig:2}, with the sliding window starting at sites 5 and 7. The second term involving $g_7g_8$ yields an operator 
$
    (\prod_{\mathrm{left}}G)
    Z_6Y_7Y_8Z_9
    (\prod_{\mathrm{right}}G)
$,
which can be measured with a settings of the type indicated by $M_2$ in Fig.~\ref{fig:2}. 

To measure term (c), we expand the corresponding operator of Fig.~\ref{fig:SI1} as
\begin{equation}
\begin{aligned}
\label{eq:four_gs}
    G_5
    E_6
    E_7
    G_8
    &\propto
    \underbrace{
    \Big{(}
    1
    +
    g_5g_8
    +
    g_5
    +
    g_8
    -
    g_6
    -
    g_7
    \Big{)}}_{M_1}
    -
    \underbrace{
    \Big{(}
    g_5g_6
    +
    g_7g_8
    +
    g_5g_7g_8
    +
    g_5g_6g_8
    \Big{)}}_{M_2}
    \\&+
    \underbrace{
    g_5g_6g_7g_8
    }_{M_3}
    +
    \underbrace{
    g_5g_6g_7
    }_{M_4}
    +
    \underbrace{
    g_6g_7g_8
    }_{M_5}
    +
    \underbrace{
    g_6g_7
    }_{M_6},
\end{aligned}
\end{equation}
where we have omitted products of operators $G$ to the left and right of $G_5E_6E_7G_8$ for brevity. The braces and labels below each term correspond to the types of measurement settings of Fig.~\ref{fig:2}, which are designed to measure that particular term. Summing over all positions of the sliding windows in Fig.~\ref{fig:2} and omitting unnecessary terms at the edges, we arrive at a total of $3(N-1)$ measurement settings required to measure the simplified lower bound of Eq.~\eqref{eq:refined_bound_simp}. 

Next, we outline a method for constructing measurement settings needed to determine the term (d) in Eq.~\eqref{eq:si:4th_order}, and hence the tighter lower bound of Eq.~\eqref{eq:refined_bound}. An operator corresponding to Fig.~\ref{fig:SI1}(d) can be expanded as
\begin{equation}
\begin{aligned}
    G_3E_4G_5G_6E_7G_8
    &\propto
    G_3(1-g_4)G_5G_6(1-g_7)G_8
    \propto
    \underbrace{G_3G_5G_6G_8}_{M_1,M_2}
    -
    \underbrace{G_3g_4G_5G_6G_8}_{M_1 - M_6}
    \\&-
    \underbrace{G_3G_5G_6g_7G_8}_{M_1 - M_6}
    +
    \underbrace{G_3g_4G_5G_6g_7G_8}_{M_1 - M_{22}}.
\end{aligned}
\end{equation}
The first term contains two $G_i$s in a row and can be measured with settings $M_1$ and $M_2$ of Fig.~\ref{fig:2}. The second and the third terms contain four $G_i$s in a row and can be measured similarly to the operator of Eq.~\eqref{eq:four_gs}. To measure the last term, one can again write $G_i = (1+g_i)/2$, where $i = [3,5,6,8]$. This results in $2^4= 16$ additional operators that can not be measured simultaneously. We provide the additional measurement settings $M_7-M_{22}$ required to measure these operators in Fig.~\ref{fig:SI9}. Since each measurement setting has to be measured at $N/2$ locations, the number of additional measurement settings is $8N$ (not accounting for corrections near the edge of the cluster state). Hence, one needs to use $11N$ settings to measure the lower-bound fidelity of Eq.~\eqref{eq:refined_bound}.

In realistic experiments term (d) of Eq.~\eqref{eq:si:4th_order} and Fig.~\ref{fig:SI1} only corresponds to a small subset of the local errors. Assume that all single-qubit errors on each qubit and two-qubit errors on each neighboring pair occur with an equal probability. There are then three possible single-qubit errors and nine possible two-qubit errors, corresponding to twelve possible error patterns. The error pattern of Fig.~\ref{fig:SI1}(d) corresponds to  $Y_iY_{i+1}$ errors, where $i \in \mathrm{odd}$. Hence, the term (d) is non-zero only for $1/18$ of all local two-qubit errors~($1/24$ of all experimentally important errors), and omitting this term will most likely have little effect on the lower-bound fidelity. This agrees well with Fig.~\ref{fig:3}, which shows the comparison between the exact fidelity, the lower bound of Eq.~\eqref{eq:refined_bound} and the simplified lower bound of Eq.~\eqref{eq:refined_bound_simp}. We therefore believe that the simplified lower bound of Eq.~\eqref{eq:refined_bound_simp} will be a desirable experimental choice for a lower-bound fidelity measure. We note in particular that if the problematic $Y_iY_{i+1}$ error is the dominant error source, it could be desirable to change basis for the cluster state from $X$ to $Y$ in which case the effect of this error would be captured by patterns similar to Fig.~\ref{fig:SI1}~(a), which is much easier to measure. 

\section{Fidelity calculation, Monte Carlo simulations}\label{sec:calculation}

By construction, the fidelity measure of Eq.~\eqref{eq:refined_bound} is identical to the full fidelity for the first-order single-qubit and local two-qubit errors. To see how our fidelity measure performs under higher-order errors, we consider a simple model with Pauli error applied to each qubit with a probability $p$. Figure~\ref{fig:3} shows the results of Monte Carlo simulations where we randomly apply single qubit errors to an ideal cluster state with a fixed probability $p$. In these simulations we start by initially assuming an ideal cluster state, which corresponds to a vector of stabilizers $S = [1_1, 1_2, ..., 1_N]$. We then flip each qubit $i = [1,N]$ by applying one of the three Pauli operators with an equal probability $p/3$. Each of these errors affect the stabilizers in a different way. Pauli-$Z$ error flips the $i$th stabilizer between $1$ and $0$. Pauli-$X$ and $Y$ errors flip stabilizers $i-1,i+1$ and $i-1,i,i+1$, respectively. This procedure returns an updated vector of stabilizers $S = [S_1, S_2, ..., S_N]$. Repeating the procedure $M$ times, we calculate the fidelity as
\begin{equation}
\label{eq:Fmc}
    \mathcal{F}
    =
    \frac{1}{M}\sum_{m=1}^M \prod_{n=1}^N S_n^{(m)}.
\end{equation}
The lower-bound fidelities of Eqs.~\eqref{eq:refined_bound} and \eqref{eq:refined_bound_simp} are calculated directly from the Monte Carlo simulations and shown in Fig.~\ref{fig:3}. We apply the same Monte Carlo method to derive the fidelity of a two-dimensional cluster state and its lower bound~[Eq.~\eqref{eq:2D_refined}] shown in Fig.~3(c,d) of the main text.

\section{Analytical solution, 1D cluster state}\label{sec:analytical}
For completeness, below we derive analytical expressions for the full fidelity and the lower-bound fidelities of Eqs.~\eqref{eq:refined_bound}, \eqref{eq:refined_bound_simp} for the case of one-dimensional cluster states affected by single-qubit errors. 

The full fidelity $\mathcal{F} = \prod_i G_i$ is non-zero when all stabilizers are correct. This is the case for an error-free state occurring with a probability $(1-p)^N$. Single- and two-qubit errors in our simple model will flip at least one stabilizer and therefore such state will not contribute to the fidelity. The correct state containing no-errors can be realized by three errors in adjacent qubits $Z_{i-1}X_{i}Z_{i+1}$, i.e., this error is one of the  stabilizer operators and hence do not affect the state.  This possibility occurs with a probability $N(p/3)^3$. For states with few qubits this quantity is small compared to lower-order terms in $(1-p)^N$ unless there is a sizable probability of errors, in which case the fidelity is very limited. For large $N$, it is small compared to the third-order term in the expansion of $(1-p)^N$, 
\begin{equation}
Np^3 \ll \begin{pmatrix}N \\ 3 \end{pmatrix}p^3 ~\mathrm{for}~ N \gg 1.
\end{equation}
Therefore this probability of multiple errors accidentally yielding the correct state can be ignored and the full fidelity is well approximated by
\begin{equation}\label{eq:Fan}
    \mathcal{F}_{\mathrm{an}}
    \approx
    (1-p)^N.
\end{equation}

Next, we calculate the simple two-measurement lower-bound fidelity, i.e. the expectation value of the operator $\hat{P}_0$ in Eq.~\eqref{eq:proj_naive}. To lowest order, the observable $\langle G_{\mathrm{o}} \rangle$ is non-zero when no $Z$ or $Y$ errors are present in odd qubits and no $X$ or $Y$ errors are present in even qubits, and vice versa for $\langle G_{\mathrm{e}} \rangle$. Such states appears with a probability $(1-2p/3)^{N}$. The simple lower-bound fidelity of Eq.~\eqref{eq:proj_naive} then reads
\begin{equation}
    {\langle}
    \hat{P}_0
    {\rangle}
    \approx
    2(1-\frac{2p}{3})^{N}
    -1.
\end{equation}
To first order in $p$, this lower-bound fidelity is $Np/3$ smaller than the true fidelity due to double counting of Pauli-$Y$ errors, as we discus in the main text.

\begin{figure}[h!]
\centering
\includegraphics[width=0.5\columnwidth]{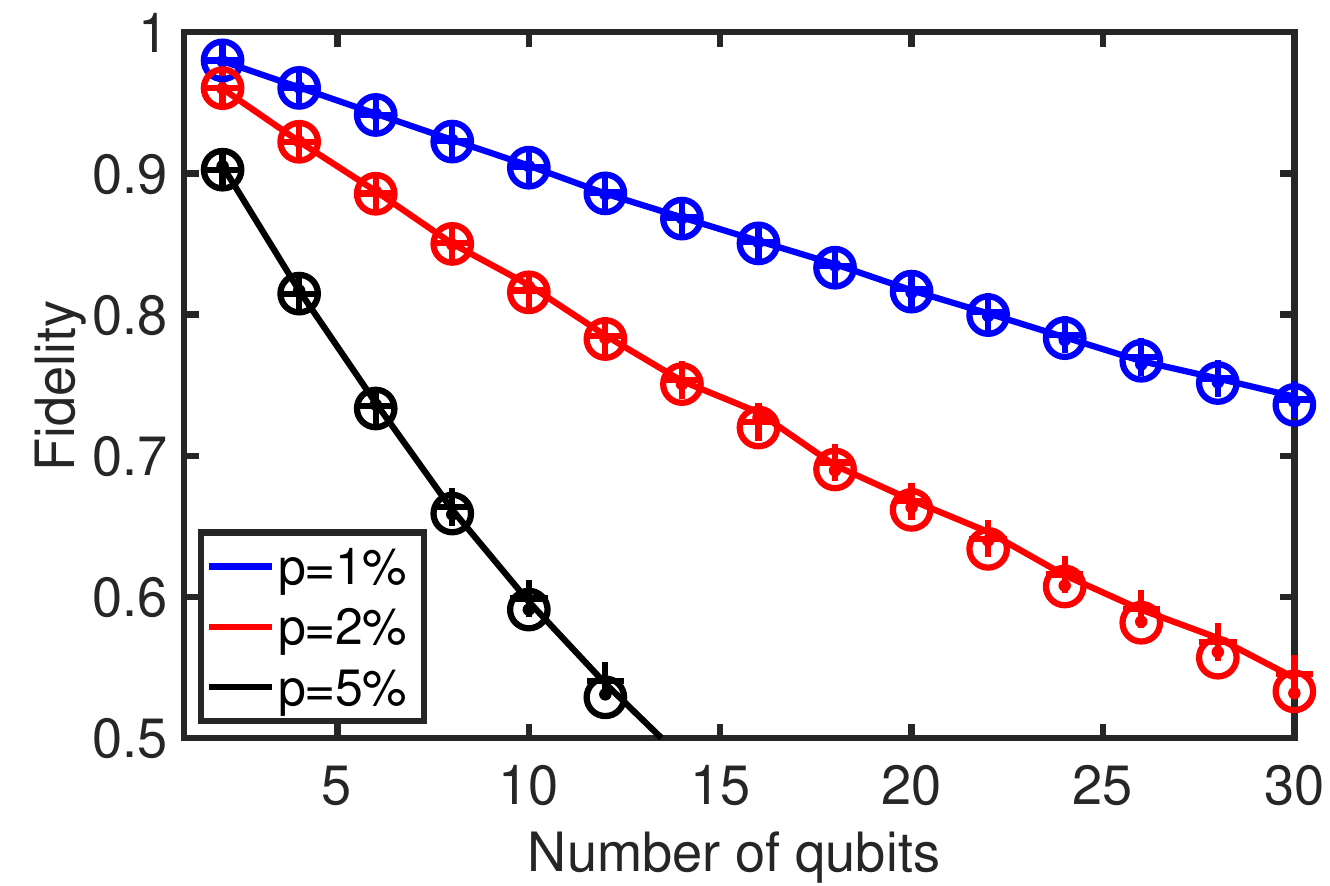}
\caption{\label{fig:SI4} Validity of the analytical solution. Solid lines and dots correspond to the full fidelity $\mathcal{F}$~[Eq.~\eqref{eq:Fmc}] and the lower-bound fidelity $\langle\hat{P}_{1D}\rangle$ obtained from Monte Carlo simulations for different single-qubit error probabilities $p$. Crosses and circles correspond, respectively, to the analytically calculated full fidelity $\mathcal{F}_{\mathrm{an}}$~[Eq.~\eqref{eq:Fan}] and the lower-bound fidelity $\langle\hat{P}_{1D}\rangle$.}
\end{figure}

To calculate the refined lower bound of Eq.~\eqref{eq:refined_bound}, one needs to evaluate the expectation value of the operator~\eqref{eq:si:4th_order},
\begin{equation}\label{eq:si:4th_order_2}
    \begin{aligned}
    \Big{\langle}
    \sum_{\substack{i \in \mathrm{odd} \\ j \in \mathrm{even} \\ j\geq i-3}}
    E_i
    E_j
    \prod_{\substack{k\in \mathrm{odd} \\ k<i}}
    G_{k}
    \prod_{\substack{m \in \mathrm{even} \\ m>j}}
    G_{m}\Big{\rangle}
    &=
    \Big{\langle}{\sum_{\substack{i \in \mathrm{odd} \\ j \in \mathrm{even} \\ j>i+1}}
    E_i
    E_j
    \prod_{\substack{k\in \mathrm{odd} \\ k<i}}
    G_{k}
    \prod_{\substack{m \in \mathrm{even} \\ m>j}}
    G_{m}\Big{\rangle}
    +
    \Big{\langle}
    \sum_{i \in \mathrm{odd}}
    E_i
    E_{i+1}
    \prod_{\substack{k\in \mathrm{odd} \\ k<i}}
    G_{k}
    \prod_{\substack{m \in \mathrm{even} \\ m>i+1}}
    G_{m}}\Big{\rangle}
    \\&+
    \Big{\langle}{\sum_{i \in \mathrm{odd}}
    E_i
    E_{i-1}
    \prod_{\substack{k\in \mathrm{odd} \\ k<i}}
    G_{k}
    \prod_{\substack{m \in \mathrm{even} \\ m>i-1}}
    G_{m}}\Big{\rangle}
    +
    \Big{\langle}{\sum_{i \in \mathrm{odd}}
    E_i
    E_{i-3}
    \prod_{\substack{k\in \mathrm{odd} \\ k<i}}
    G_{k}
    \prod_{\substack{m \in \mathrm{even} \\ m>i-3}}
    G_{m}}\Big{\rangle}.
    \end{aligned}
\end{equation}
To  second order in $p$, each of the four terms reads
\begin{equation}\label{eq:ee0}
    \begin{aligned}
    \Big{\langle}
    \sum_{\substack{i \in \mathrm{odd} \\ j \in \mathrm{even} \\ j>i+1}}
    E_iE_{j}
    \prod_{\substack{k\in \mathrm{odd} \\ k<i}}
    G_k
    \prod_{\substack{m\in \mathrm{even} \\ m>j}}
    G_m
    \Big{\rangle}
    &=
    \sum_{i \in \mathrm{even}}
    \sum_{\substack{j \in \mathrm{odd} \\ j>i+2}}
    \Big{(}
    \frac{4p}{3}
    \Big{)}^2
    \Big{(}
    1-\frac{2p}{3}
    \Big{)}^{N+i-j}
\end{aligned}
\end{equation}
\begin{equation}\label{eq:ee1}
    \begin{aligned}
    \Big{\langle}
    \sum_{i \in \mathrm{odd}}
    E_iE_{i+1}
    \prod_{\substack{k\in \mathrm{odd} \\ k<i}}
    G_k
    \prod_{\substack{m\in \mathrm{even} \\ m>i+1}}
    G_m
    \Big{\rangle}
    &=
    \frac{Np}{3}
    \Big{(}
    1-\frac{2p}{3}
    \Big{)}^{N-3}
    +
    \frac{N}{2}
    \Big{(}
    \frac{p}{3}
    \Big{)}^2
    \Big{(}
    1-\frac{2p}{3}
    \Big{)}^{N-2}
\end{aligned}
\end{equation}

\begin{equation}
    \begin{aligned}
    \Big{\langle}
    \sum_{i \in \mathrm{odd}}
    E_iE_{i-1}
    \prod_{\substack{k\in \mathrm{odd} \\ k<i}}
    G_k
    \prod_{\substack{m\in \mathrm{even} \\ m>i-1}}
    G_m
    \Big{\rangle}
    &=
    \frac{N}{2}
    \Big{(}
    \frac{p}{3}
    \Big{)}^2
    \Big{(}
    1-\frac{2p}{3}
    \Big{)}^{N-2}
    +
    N
    \Big{(}
    \frac{p}{3}
    \Big{)}^2
    \Big{(}
    1-\frac{2p}{3}
    \Big{)}^{N-3}
    \Big{(}1-p\Big{)}
\end{aligned}
\end{equation}
\begin{equation}\label{eq:ee3}
    \begin{aligned}
    \Big{\langle}
    \sum_{i \in \mathrm{odd}}
    E_iE_{i-3}
    \prod_{\substack{k\in \mathrm{odd} \\ k<i}}
    G_k
    \prod_{\substack{m\in \mathrm{even} \\ m>i-3}}
    G_m
    \Big{\rangle}
    &=
     N
    \Big{(}
    \frac{p}{3}
    \Big{)}^2
    \Big{(}
    1-\frac{2p}{3}
    \Big{)}^{N-4}
    \Big{(}1-p\Big{)}^2
\end{aligned}
\end{equation}
Substituting Eqs.~\eqref{eq:si:4th_order_2}--\eqref{eq:ee3} into
\begin{equation}
\label{eq:P1D}
    \langle
    \hat{P}_{1D}
    \rangle
    =
    \langle
    \hat{P}_{0}
    \rangle
    +
    \Big{\langle}
    \sum_{\substack{i \in \mathrm{odd} \\ j \in \mathrm{even} \\ j \geq i-3}}
    E_i
    E_j
    \prod_{\substack{k\in \mathrm{odd} \\ k<i}}
    G_{k}
    \prod_{\substack{m \in \mathrm{even} \\ m>j}}
    G_{m}
    \Big{\rangle},
\end{equation}
we arrive at the analytical lower-bound fidelity shown in Fig.~\ref{fig:SI4}. As seen from the figure, the simple analytical solution derived above matches well with the Monte Carlo solution.

\begin{figure}[t]
\centering
\includegraphics[width=0.75\columnwidth]{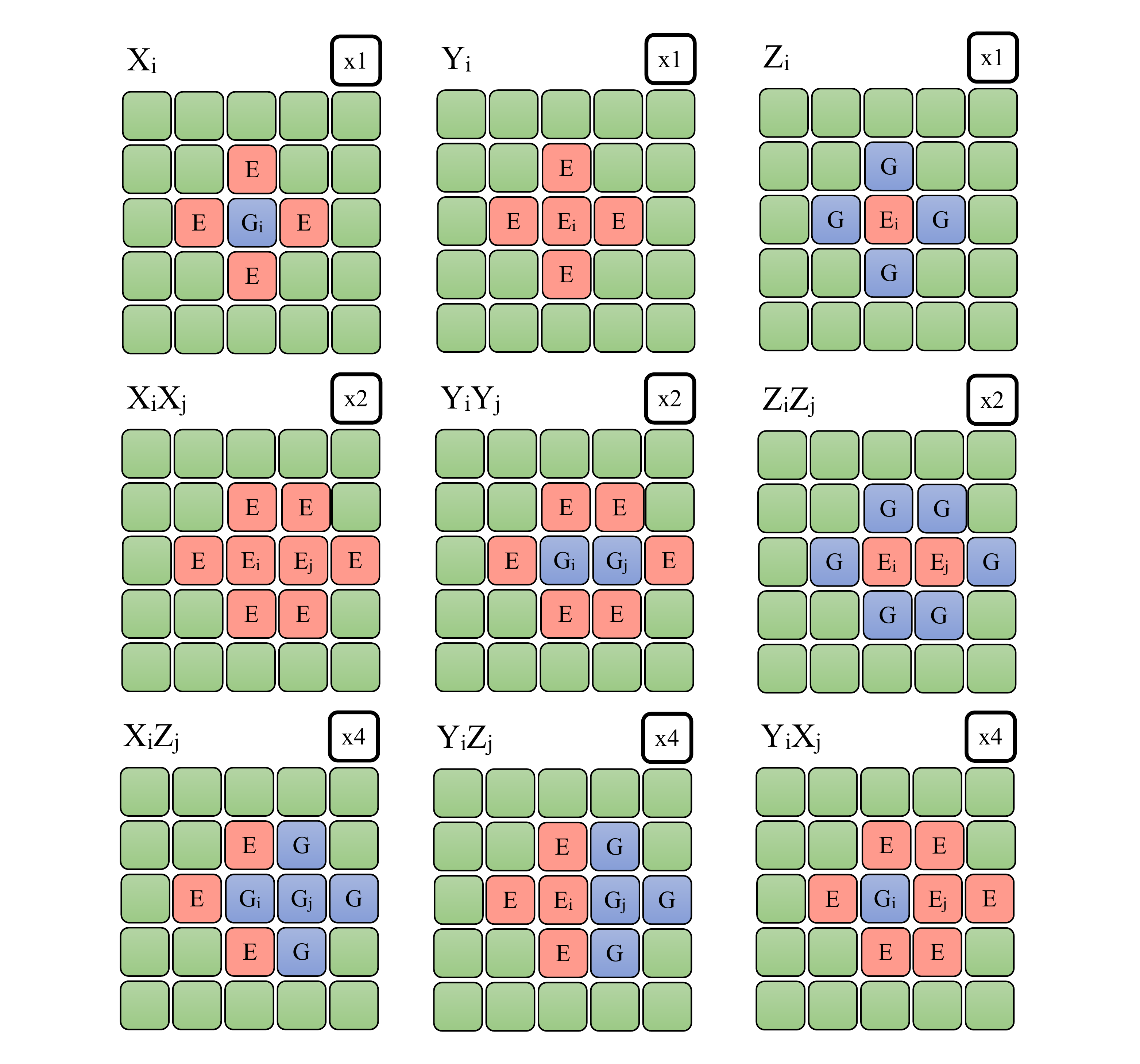}
\caption{\label{fig:SI6} Stabilizer error patterns corresponding to single Pauli errors in qubit $i$ and pairs of Pauli errors in nearest-neighboring qubits $i$ and $j$. Salmon and blue squares correspond to flipped and correct stabilizers, respectively. Green squares are not affected by the errors. The number in the top-right corner of each pattern shows the number of possible orientations each error can have. Taking all orientations into account, there are 18 possible two-qubit error patters.}
\end{figure}

\section{Number of measurement settings, 2D cluster state}\label{sec:2D_states}

Before introducing the measurement settings for the case of two-dimensional cluster states, we note that there is a slight difference in the lower-bound fidelity expressions for odd and even squared-shaped clusters. Formula~\eqref{eq:2D_refined} of the main text as well as the analysis below applies to the clusters with odd number of qubits along each direction. In the Appendix~\ref{app:even2d}, we provide a lower-bound fidelity expression tailored for even clusters.

As in the case of one-dimensional cluster states, the experimentally important errors described by $(1-G_{\mathrm{o}})(1-G_{\mathrm{e}})$ include single-qubit and local two-qubit errors shown in Fig.~\ref{fig:SI6}. The lower-bound fidelity can be obtained from the full fidelity of Eq.~\eqref{eq:projector_g_split} by keeping only the terms that correspond to such experimentally relevant errors, yielding
\begin{equation}\label{eq:refined_bound_2D_SI}
    \begin{aligned}
    \hat{P}_{2D}
    &=
    G_{\mathrm{e}}
    +
    G_{\mathrm{o}}
    -1
    +
    \sum_{i \in \mathrm{odd}}
    \sum_{\substack{j \in \mathrm{even} \\ j-i > N_x}}
    E_i
    E_j
    \prod_{\substack{k\in \mathrm{odd} \\ k<i}}
    G_{k}
    \prod_{\substack{m \in \mathrm{even} \\ m>j}}
    G_{m}
    +
    \sum_{i \in \mathrm{odd}}\sum_{\substack{j \in \mathrm{even} \\ j \in \mathcal{N}_i}}
    E_i
    E_j
    \prod_{\substack{k\in \mathrm{odd} \\ k<i}}
    G_{k}
    \prod_{\substack{m \in \mathrm{even} \\ m>j}}
    G_{m}
    \\&=
    G_{\mathrm{e}}
    +
    G_{\mathrm{o}}
    -1
    +
    \underbrace{\sum_{i \in \mathrm{odd}}
    \sum_{\substack{j \in \mathrm{even} \\ j-i > N_x}}
    E_i
    E_j
    \prod_{\substack{k\in \mathrm{odd} \\ k<i}}
    G_{k}
    \prod_{\substack{m \in \mathrm{even} \\ m>j}}
    G_{m}}_{(a)}
    +
    \underbrace{\sum_{i \in \mathrm{odd}}
    E_i
    E_{i+N_x}
    \prod_{\substack{k\in \mathrm{odd} \\ k<i}}
    G_{k}
    \prod_{\substack{m \in \mathrm{even} \\ m>i+N_x}}
    G_{m}}_{(b)}
    \\&+
    \underbrace{\sum_{i \in \mathrm{odd}}
    E_i
    E_{i+1}
    \prod_{\substack{k\in \mathrm{odd} \\ k<i}}
    G_{k}
    \prod_{\substack{m \in \mathrm{even} \\ m>i+1}}
    G_{m}}_{(c)}
    +
    \underbrace{\sum_{i \in \mathrm{odd}}
    E_i
    E_{i-1}
    \prod_{\substack{k\in \mathrm{odd} \\ k<i}}
    G_{k}
    \prod_{\substack{m \in \mathrm{even} \\ m>i-1}}
    G_{m}}_{(d)}
    +
    \underbrace{\sum_{i \in \mathrm{odd}}
    E_i
    E_{i-N_x}
    \prod_{\substack{k\in \mathrm{odd} \\ k<i}}
    G_{k}
    \prod_{\substack{m \in \mathrm{even} \\ m>i-N_x}}
    G_{m}}_{(e)},
    \end{aligned}
\end{equation}
where $N_x\sim \sqrt{N}$ is the horizontal size of the cluster and $\mathcal{N}_i = \{ i-1,i+1,i-N_x,i+N_x \}$ is a set of all nearest neighbors of qubit $i$. 

\begin{figure}[h!]
\centering
\includegraphics[width=0.95\columnwidth]{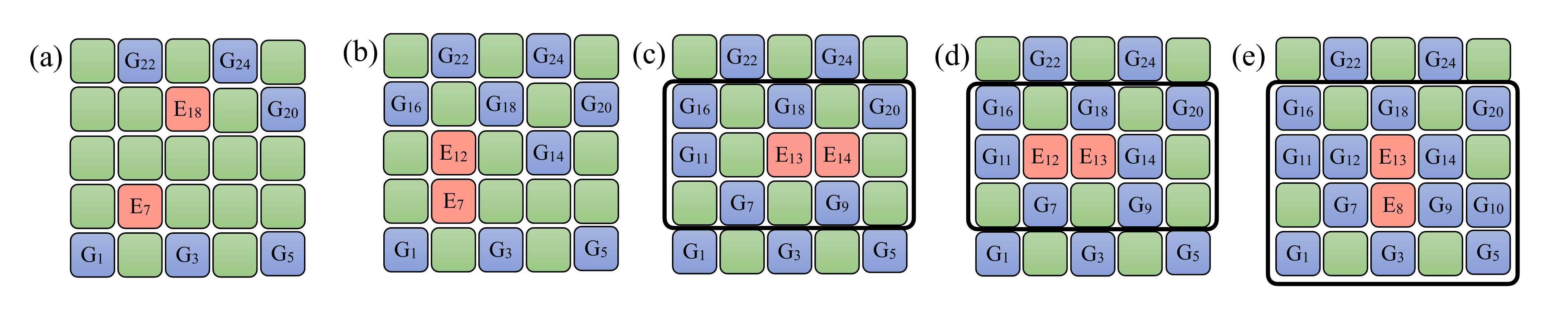}
\caption{\label{fig:SI7} Pictorial representation of the terms (a)--(e) of Eq.~\eqref{eq:refined_bound_2D_SI}. The operators of panels (a) and (b) can be measured with, respectively, one and two measurement settings, while panels (c)--(e) require a large number of measurement settings growing exponentially with $N_x$ inside the framed regions.}
\end{figure}

Terms (a)--(e) of Eq.~\eqref{eq:refined_bound_2D_SI} are shown in Fig.~\ref{fig:SI7}. The complexity of the measurement procedure increases from panel (a) to panel (d): The number of measurement setting increases exponentially with the number of neighboring operators. The operators of panels (a) and (b) have zero or one neighboring operators. This means that they can be measured with single measurement setting for each location of the first error. This means that the measurements of these terms require a number of settings linear in $N$. In comparison, the number of settings required for measuring the operators of panels (c)--(e) scales approximately exponentially with the horizontal size $N_x$ of the cluster since these have an increasing number of neighboring operators. On the other hand, the probability of the errors detectable with the measurement settings of Fig.~\ref{fig:SI7} decreases from panel (a) to panel (e). Therefore, one can expect only a small measurement error if  terms (c)--(e) in Eq.~\eqref{eq:refined_bound_2D_SI} are ignored.  This is analogous to ignoring  term (d) of Eq.~\eqref{eq:si:4th_order} in the one-dimensional case. Omitting these terms yields the lower-bound fidelity of Eq.~\eqref{eq:2D_refined} of the main text,
\begin{equation}\label{eq:2D_refined_app}
    \hat{P}_{2D}
    =
    G_{\mathrm{e}}
    +
    G_{\mathrm{o}}
    -1
    +
    \sum_{i \in \mathrm{odd}}
    \sum_{\substack{j \in \mathrm{even} \\ j-i \geq N_x}}
    E_i
    E_{j}
    \prod_{\substack{k\in \mathrm{odd} \\ k<i}}
    G_{k}
    \prod_{\substack{m \in \mathrm{even} \\ m>j}}
    G_{m}.
\end{equation}

\begin{figure}[h!]
\centering
\includegraphics[width=0.6\columnwidth]{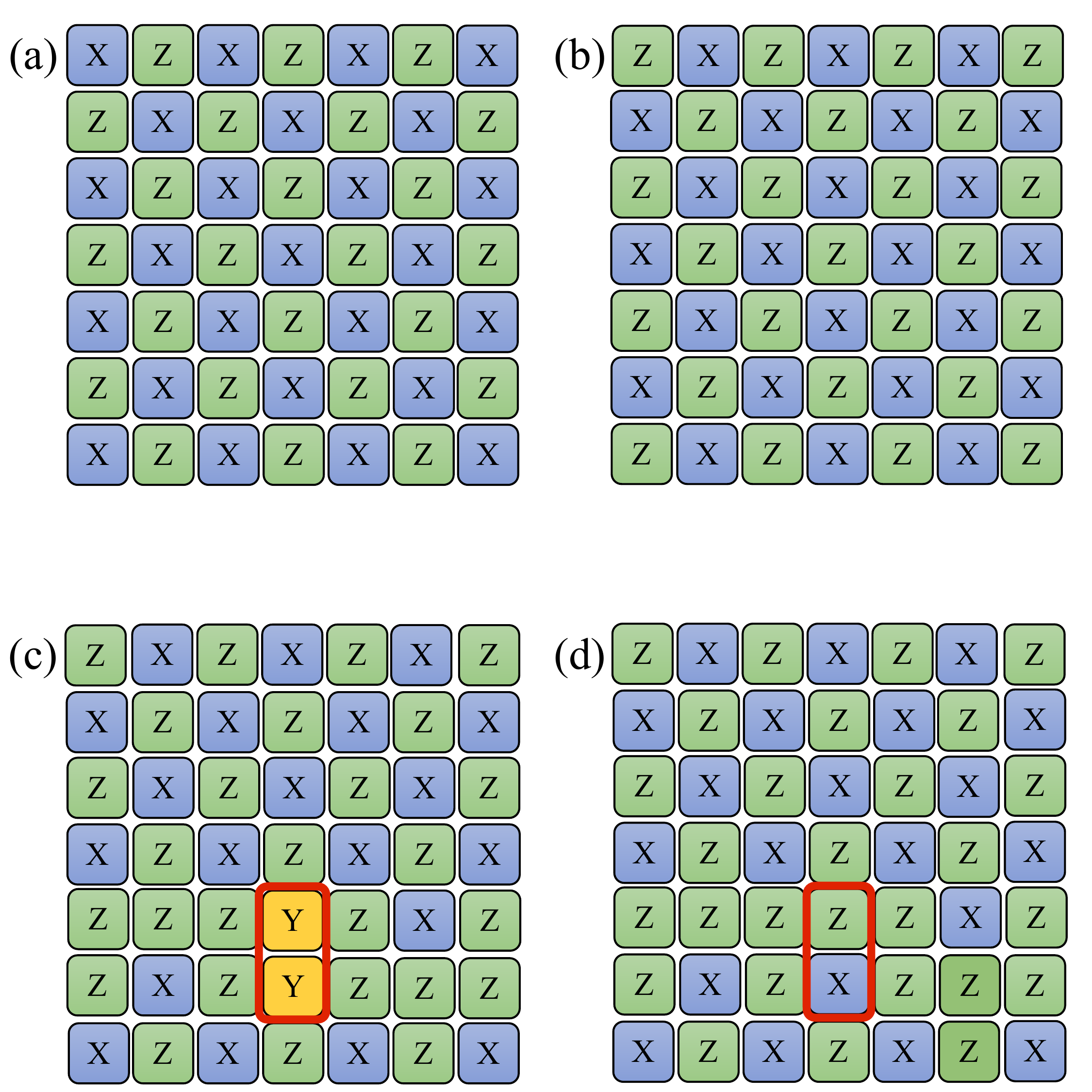}
\caption{\label{fig:SI8} Measurement settings required to determine the expectation value of the lower-bound fidelity~\eqref{eq:2D_refined_app} of a two-dimensional cluster state. The measurement settings of panels (a) and (b) are used to measure the operators $G_{\mathrm{o}}$ and $G_{\mathrm{e}}$, respectively. 
Additionally, the measurement settings of panels (c) and (d) are used to measure the last term of Eq.~\eqref{eq:2D_refined_app}, i.e. the terms (a) and (b) of Eq.~\eqref{eq:refined_bound_2D_SI}. Here, the red window is sliding with step 2. Qubits to the left and right from the sliding window are measured as shown in the figure. Qubits in the first row to the left and in the second row to the right from the red rectangle are measured in the $Z$ basis. Odd~(even) qubits in the first row to the right from the red rectangle are measured in the $Z$~($X$) basis. Odd~(even) qubits in the second row to the left from the red rectangle are measured in the $X$~($Z$) basis. 
Odd and even qubits below (above) the sliding window are measured in the $X$ and $Z$ ($Z$ and $X$) bases, respectively. The sliding window takes $N_x/2$ positions on $N_y-1$ rows. Hence, the total number of measurement settings is $N_x(N_y-1)+2 \approx N$.} 
\end{figure}

To measure the first two terms of the operator above, one needs the two measurement settings shown in Fig.~\ref{fig:SI8} (a,b). One also needs to measure the last term, which corresponds to the terms (a) and (b) of Eq.~\eqref{eq:refined_bound_2D_SI}. First, we note that by measuring  term (b) one automatically measures the term (a), which can be seen by comparing panels (a) and (b) of Fig.~\ref{fig:SI7}. Next, we expand the operators corresponding to  two flipped stabilizers in term (b) of Eq.~\eqref{eq:refined_bound_2D_SI} as $E_iE_{i+N_x} = (1 - g_i - g_{i+N_x} + g_ig_{i+N_x})/4$. The operator
\begin{equation}
    g_ig_{i+N_x}
    \prod_{\substack{k\in \mathrm{odd} \\ k<i}}
    G_{k}
    \prod_{\substack{m \in \mathrm{even} \\ m>i+N_x}}
    G_{m}
\end{equation}
corresponds to the pattern of Fig.~\ref{fig:SI8}(c). Since $i$ here takes odd values, this results in 
$N_x(N_x-1)/2 \approx N/2$ 
measurement settings. Analogously, the operators
\begin{equation}
    g_i
    \prod_{\substack{k\in \mathrm{odd} \\ k<i}}
    G_{k}
    \prod_{\substack{m \in \mathrm{even} \\ m>i+N_x}}
    G_{m}
\end{equation}
correspond to the pattern of Fig.~\ref{fig:SI8}(d) that can have another $N_x  (N_x-1)/2 \approx N/2$ configurations. The total number of measurement settings required to determine the lower-bound fidelity of Eq.~\eqref{eq:2D_refined_app} is therefore $N_x(N_x-1)+2 \approx N$.

Now we analyse the performance of the lower-bound fidelity~\eqref{eq:2D_refined_app}. For single-qubit errors, the expression~\eqref{eq:2D_refined_app} yields the same fidelity as the full projector of Eq.~\eqref{eq:projector_g_split} to lowest order in the error. Therefore, the operator~\eqref{eq:2D_refined_app} is a good lower-bound fidelity measure in the presence of single-qubit errors, as shown in Fig.~\ref{fig:3} of the main text. 

To estimate the performance of the fidelity measure~\eqref{eq:2D_refined_app} under local two-qubit errors, we assume again a simple error model such that
\begin{equation}
    P(\sigma^{i}_l \sigma^{j}_m) = p/9,
\end{equation}
where $i,j$ denote the nearest-neighbor qubits and $l,m \in \{ x,y,z \}$. Taking into account different orientations of errors in nearest-neighbor qubits, there are 18 possible configurations shown in Fig.~\ref{fig:SI6}. Comparing the error patterns of Fig.~\ref{fig:SI6} with the  operators of Fig.~\ref{fig:SI7}, one can see that the operators of Fig.~\ref{fig:SI7}(e) detects 1/36 of the possible errors, the  of Fig.~\ref{fig:SI7}(d) detects 1/36 of the possible errors, and the operators of Fig.~\ref{fig:SI7}(c) detects 1/9 of possible errors. Therefore, the lower-bound fidelity of Eq.~\eqref{eq:2D_refined_app} correctly detects $15/18 \approx 85\%$ of all local two-qubit errors, which agrees well with the numerically-simulated fidelities of Fig.~3(d) of the main text. 

\section{Even 2D cluster states}
\label{app:even2d}

\begin{figure}[h!]
\centering
\includegraphics[width=0.6\columnwidth]{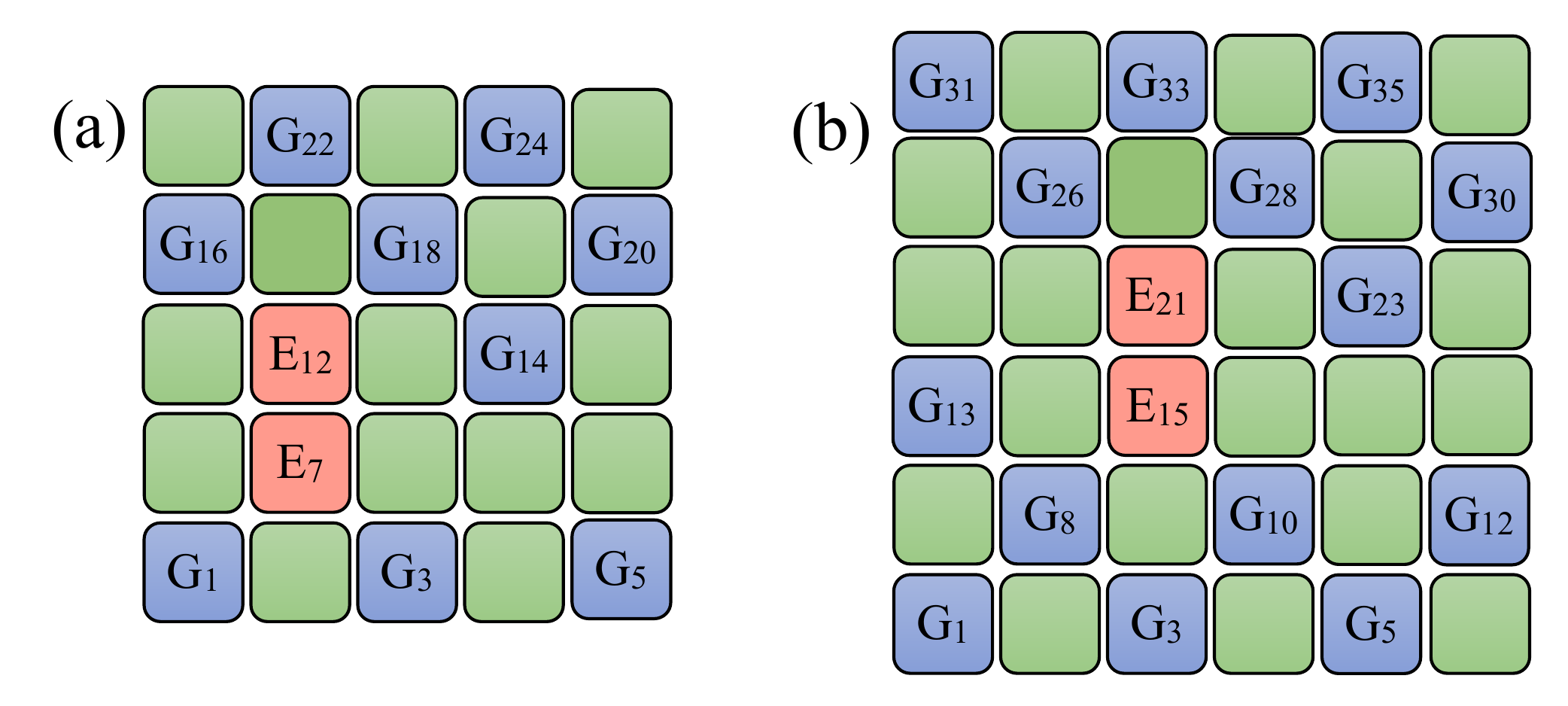}
\caption{\label{fig:SI10} Labeling (a)~odd and (b)~even graphs. Panels (a) depicts term (b) of Eq.~\eqref{eq:refined_bound_2D_SI} for an odd cluster. Panels (b) depicts the corresponding term for an even cluster, i.e. Eq.~\eqref{eq:refined_bound_2D_SI_even}.}
\end{figure}

In order to tailor the lower-bound fidelity to even clusters, we modify Eq.~\eqref{eq:2D_refined} of the main text as follows:
\begin{equation}\label{eq:refined_bound_2D_SI_even}
    \hat{P}_{2D}^{(\textrm{even})}
    =
    G_{\mathrm{e}}
    +
    G_{\mathrm{o}}
    -1
    +
    \sum_{i \in S_1}
    \sum_{\substack{j \in S_2 \\ j-i \geq N_x}}
    E_i
    E_{j}
    \prod_{\substack{k\in S_1 \\ k<i}}
    G_{k}
    \prod_{\substack{m \in S_2 \\ m>j}}
    G_{m},
\end{equation}
where the set $S_1$ contains odd labels on the odd rows and even labels on the even rows, while the set $S_2$ contains even labels on the odd rows and odd labels on the evens rows. We note that the same modification has to be applied to the operators $G_{\mathrm{e}}$ and $G_{\mathrm{o}}$ above, i.e.
\begin{equation}
    \begin{aligned}
    G_{\mathrm{o}} &= \prod_{i \in \mathrm{S_1}}G_i
    \\
    G_{\mathrm{e}} &= \prod_{i \in \mathrm{S_2}}G_i.
    \end{aligned}
\end{equation}
Such modification allows to correctly label the stabilizers we want to measure when going from odd to even clusters, i.e. from Fig.~\ref{fig:SI10}(a) to Fig.~\ref{fig:SI10}(b).

\end{document}